\documentclass{statsoc}

\usepackage{amsmath}
\usepackage{amssymb}
\usepackage{bm}
\usepackage{epstopdf}
\usepackage{color}
\usepackage{comment}
\usepackage[letterpaper, margin = 1in]{geometry}
\usepackage[textwidth=8em,textsize=small]{todonotes}

\usepackage{etoolbox}

\makeatletter
\patchcmd{\@makecaption}
  {\parbox}
  {\advance\@tempdima-\fontdimen2} 
  {}{}
\makeatother  

\usepackage{graphicx}
\usepackage[caption = false]{subfig}

\def\bbeta{\bm \beta}
\def\btheta{\bm \theta}

\def\boldeta{\bm \eta}
\def\eigen{\mbox{eigen}}

\newtheorem{theorem}{Theorem}[section]
\newtheorem{lemma}{Lemma}[section]

\newtheorem{remark}{Remark}[section]

\def\spacingset#1{\renewcommand{\baselinestretch}%
{#1}\small\normalsize} \spacingset{1}

\title[Scalable Sparse Cox's Regression]{Scalable Sparse Cox's Regression for Large-Scale Survival Data via Broken Adaptive Ridge }

\author{Eric S. Kawaguchi}
\address{University of California, Los Angeles,
USA.} 

\author{Marc A. Suchard}
\address{University of California, Los Angeles,
USA.} 

\author{Zhenqiu Liu}
\address{Cedars-Sinai Medical Center, Los Angeles,
USA.} 

\author[E.S. Kawaguchi, M.A. Suchard, Z. Liu, and G. Li]{and Gang Li}
\address{University of California, Los Angeles,
USA.} 
\email{vli@ucla.edu}

\begin{document}
\begin{abstract}
This paper develops a new scalable sparse Cox regression tool for  sparse high-dimensional 
massive sample size  (sHDMSS) 
survival data. The method is  a local $L_0$-penalized Cox regression via repeatedly performing reweighted $L_2$-penalized Cox regression.
We show that the resulting estimator enjoys the best of $L_0$- and $L_2$-penalized Cox regressions while overcoming their limitations. Specifically, the estimator is selection consistent,  oracle for parameter estimation, and  possesses
a grouping property for highly correlated covariates. Simulation results suggest that when the sample size is large, the proposed method with pre-specified tuning parameters has a comparable or better performance than some popular penalized regression methods. More importantly, because
the method naturally enables adaptation of efficient algorithms for massive $L_2$-penalized optimization and does not require costly data driven
tuning parameter selection, it has a significant computational advantage for sHDMSS data, offering an average of 5-fold speedup over its closest competitor in empirical studies. 

\end{abstract}


\section{Introduction}
\label{s1:intro}
Advancements in medical informatics tools and high-throughput biological experimentation are making large-scale data routinely
accessible to researchers, administrators, and policy-makers.  This data deluge poses new challenges and critical barriers for quantitative researchers as existing statistical methods and software grind to a halt when analyzing these large-scale datasets, and
calls for appropriate methods that can readily fit large-scale data.  This paper primarily concerns survival analysis of 
{\em sparse high-dimensional}  {\em massive sample size} (sHDMSS) data, a particular type of large-scale data with the following characteristics:  
1) high-dimensional  with a large number  of covariates ($p_n$ in thousands or tens of thousands), 2) massive in sample-size
($n$ in thousands to hundreds of millions), 3) sparse in covariates with only a very small portion of covariates 
being nonzero for each subject, and 4) rare in event rate.  A typical example of sHDMSS data is
the pediatric trauma mortality data  \citep{mittal2013high} from the National Trauma Databank  (NTDB) maintained by the American College of Surgeons \citep{mittal2013high}. This data set includes 210,555 patient records of injured children under 15 collected over 5 years from 2006 -2010. Each patient record includes 125,952 binary covariates that indicate the presence, or absence, of an attribute (ICD9 Codes, AIS codes, etc.) as well as their two-way interactions. The data matrix is extremely sparse with less than 1\% of the covariates being non zero. The event rate is also very low at $2\%$.   Another application domain where sHDMSS data are common is drug safety studies 
that use massive patient-level databases such as the U.S.~FDA's Sentinel Initiative (\textcolor{blue}{\it https://www.fda.gov/safety/fdassentinelinitiative/ucm2007250.htm}) and the  Observational Health Data Sciences and Informatics (OHDSI) program 
(\textcolor{blue}{\it https://ohdsi.org/}) to study rare adverse events with hundreds of millions of patient records and tens of thousands of patient attributes that are sparse in the covariates.

sHDMSS survival data presents multiple challenges to quantitative researchers. 
First, not all of the thousands of covariates are expected to be relevant to an outcome of interest. 
Traditionally, researchers hand-pick subject characteristics to
include in an analysis. However, hand picking can introduce not only bias, but also a source of variability between
researchers and studies. 
Moreover, it would become impractical and infeasible in large-scale evidence generation when 
hundreds or thousands of analyses are to be performed  \citep{suchard2017}.  
Hence, automated  
sparse regression methods are desired. Secondly, the massive sample size presents a critical barrier to the application of existing sparse survival regression methods in a high-dimensional setting. 
While there are available many sparse survival regression methods  
\citep{tibshirani1997lasso, fan2002variable, zhang2007adaptive, zhang2010nearly, simon2011regularization, johnson2012log, su2016sparse}, 
current methods and standard software become inoperable for large datasets due to high computational costs and large memory requirements. 
\cite{mittal2013high} presented tools for fitting $L_2$ (ridge) and $L_1$ (LASSO) penalized Cox's regressions on sHDMSS data. However, it is well known that ridge regression is not sparse and that although $L_1$-penalized regression produces a sparse
solution, it tends to select too many noise variables and is biased for estimation.
Lastly,  the commonly used ``divide and conquer'' strategy for massive size data is deemed inappropriate for sHDMSS 
data  since each of the divided data would typically be too sparse for a meaningful analysis. 
Improved scalable sparse regression methods for sHDMSS data are therefore critically needed.

This paper develops a new sparse Cox regression method, named Cox broken adaptive ridge (CoxBAR) regression, 
which starts with an initial Cox ridge estimator and then iteratively performs a reweighted ridge regression that aims to approximate 
an $L_0$-penalized regression. 
It is well known that 
$L_0$-penalized regression is natural for variable selection and parameter estimation with some optimal properties \citep{akaike1974new, schwarz1978estimating, volinsky2000bayesian, shen2012likelihood}, but it is also known to have some limitations such as being unstable 
\citep{breiman1996heuristics} and not scalable to high-dimensional settings.  
The CoxBAR method aims to yield a local solution of $L_0$-penalized Cox regression
that preserves some desirable properties of $L_0$-penalized Cox regression while avoiding its limitations.
First, the CoxBAR estimator is stable and easily scalable to high dimensional covariates.
Second, the CoxBAR estimator in fact enjoys the best of $L_0$-penalized regression and the oracle ridge estimator.
We will show that the reweighted ridge regression at each iteration step shrinks the small values of the initial Cox ridge estimator towards zero and drives its large values towards an oracle ridge
estimator. Consequently, the resulting CoxBAR estimator is selection
consistent and its nonzero component behaves  like the oracle ridge estimator that is asymptotically consistent, normal, and has a grouping property for highly correlated covariates. Lastly and most importantly,  the CoxBAR method has a computational advantage over other penalized regression methods  for fitting sHDMSS survival data since it naturally takes advantage of existing efficient algorithms for massive $L_2$-penalized optimization (see Section 2.2) and  does not require costly
data-driven tuning parameter selection (see Section 2.1.4 and Section \ref{s1:sens_sim}).

The idea of iteratively reweighted penalizations dates back at least to the well-known 
Lawson's algorithm \citep{lawson1961contributions} in
classical approximation theory, which has been applied to various applications including $L_d$ ($0 < d < 1$)
minimization \citep{osborne1985finite}, sparse signal reconstruction \citep{gorodnitsky1997sparse},  compressive
sensing \citep{candes2008enhancing, chartrand2008iteratively, gasso2009recovering, daubechies2010iteratively, wipf2010iterative}, and variable selection for linear models and generalized linear models \citep{liu2016efficient, frommlet2016adaptive}.  
However, except for the linear model, current iteratively reweighted penalization algorithms are  not readily applicable to sHDMSS data. For example,
the commonly used Newton-Ralphson algorithm in each reweighted penalization becomes unsuitable for large-scale settings with large $n$ and $p_n$ due to high computational costs, high memory requirements, and numerical instability.  Furthermore, computation
of the Cox partial likelihood and its derivatives is particularly demanding for massive sample size data since the required number of operations grows  at the rate of $O(n^2)$. One of the key contributions of this paper is to 
develop an efficient implementation of CoxBAR for Cox regression with sHDMSS survival data by adapting existing efficient massive $L_2$-penalized Cox regression techniques, which include employing
a column relaxation with logistic loss (CLG) algorithm using 1D updates and a one-step 
Newton-Raphson approximation and exploiting the sparsity in the covariate structure and the Cox partial likelihood. We will also show that CoxBAR does not require costly data-driven tuning parameter selection, which turns out to be a significant computational advantage for fitting sHDMSS survival data. Another key contribution of this paper is the rigorous development of the asymptotic  properties of the CoxBAR estimator.  To this end, we point out that
previous theoretical studies of iteratively reweighted penalization methods have  focused only on numerical convergence properties and that
statistical properties of the resulting estimator remain unexplored.
Furthermore, unlike most penalized regression methods that produce a sparse solution in a single step, 
the CoxBAR method is not sparse per se at each iteration 
and  only achieves sparsity at its limit.  Consequently, our theoretical derivations for the CoxBAR estimator are quite different from
those for a single-step oracle estimator in the literature.


In Section \ref{s1:method}, we formally define the CoxBAR estimator,  state its theoretical properties for variable selection,  parameter estimation, and grouping highly correlated covariates, and describe an efficient implementation of CoxBAR for sHDMSS survival data.
As a by-product, we also discuss how to adapt CoxBAR as a 
post-screening sparse regression method for ultrahigh dimensional covariates with relatively small sample size.  Simulation studies are presented in Section \ref{s1:sim} to {demonstrate the performance of the CoxBAR estimator with both moderate and massive sample size in various low and high-dimensional settings}. A real data example including an application of CoxBAR 
on the pediatric trauma mortality data  \citep{mittal2013high} is given in Section \ref{s1:data}. Closing remarks and discussion 
are given in Section \ref{s1:disc}.  Proofs of the theoretical results and regularity conditions needed for the derivations are collected in the Online Supplementary Material. An R package for CoxBAR is available at \verb+https:github.com/OHDSI/BrokenAdaptiveRidge. +


\section{Methodology}
\label{s1:method}

\subsection{Cox's broken adaptive ridge regression and its large sample properties}\label{sect2.1}

\subsubsection{The estimator}
Suppose that one observes a random sample of right-censored survival data consisting of $n$
independent and identically distributed triplets, $\{(\tilde{T}_i, \delta_i, \mathbf{x}_i(\cdot))\}_{i = 1}^n$, 
where for subject $i$, $\tilde{T}_i = \min(T_i,C_i)$ is the observed survival time, $\delta_i = I(T_i \leq C_i)$ is the censoring indicator, 
$T_i$ is a survival time of interest, and $C_i$ is a censoring time that is conditionally independent of $T_i$ given
a $p_n$-dimensional, possibly time-dependent, covariate vector $\mathbf{x_i}(\cdot) = (x_{i1}(\cdot), \ldots, x_{ip_n}(\cdot))^T$.

Assume the  \citet{cox1972regression} proportional hazard model 
\begin{equation}
\label{eq1:coxmodel}
 h\{t|\mathbf{x}(t)\}= h_{0}(t) \exp\{\mathbf{x}(t)^T\bbeta\}, 
 \end{equation}
where $h\{t|\mathbf{x}(t)\}$ is the conditional hazard function of  $T_i$ given $\{ \mathbf{x}(u)$, $0 \leq u \leq t,\}$, $h_{0}(t)$ is an unspecified baseline hazard function, and $\bbeta = (\beta_1, \ldots, \beta_{p_n})$ is a vector of regression coefficients. Denote by  $\bbeta_1$ and $\bbeta_2$ the first $q_n$  and remaining $p_n - q_n$ components of $\bbeta$, respectively, and define $\bbeta_0 = \left(\bbeta_{01}^T,\bbeta_{02}^T\right)^T$ as the true values of $\bbeta$ where, without loss of generality, $\bbeta_{01}=(\beta_{01} \ldots,\beta_{0{q_n}})$ is a vector of $q_n$ non-zero values and $\bbeta_{02} = \boldsymbol{0}$ is a $p_n-q_n$ dimensional vector of zeros. Further technical assumptions for $\bbeta_0$ and $p_n$ are given later in condition (C6) of Section \ref{s1:oracle}. Without loss of generality, we  work on the time interval $s \in [0,1]$ as in \cite{andersen1982cox}, which can be extended to the time interval $[0, \tau]$ for $0 < \tau < \infty$ without difficulty.
Adopting the counting process notation of \cite{andersen1982cox}, the log-partial likelihood for the Cox model is defined as
\begin{equation}
\label{eq1:loglik}
l_n(\bbeta) = \sum_{i=1}^n \int_{0}^1 \bbeta^T \mathbf{x}_{i}(s)dN_{i}(s) - \int_0^1 \ln \left[ \sum_{j=1}^n Y_j(s) \exp \{\bbeta^T\mathbf{x}_{j}(s)\} \right] d\bar{N}(s),
\end{equation} 
where for subject $i$, $Y_i(s) = I(\tilde{T}_i \geq s)$ is the at-risk process  and $N_i(s) = I(\tilde{T}_i\leq s, \delta_i =1)$ is the counting process of the uncensored event  with intensity process $h_i(t|\bbeta) = h_0(t)Y_i(t)\exp\{\mathbf{x}_i(t)^T\bbeta\}$ and $\bar{N} = \sum_{i=1}^n N_i$. Let $H_i(t) = \int_0^1 h_i(u, \bbeta_0) du$, then $M_i(t) = N_i(t) - H_i(t)$ is a local square integrable martingale with respect to filtration
$\mathcal{F}_{t, i} = \sigma\{N_i(u), \mathbf{x}_i(u^+), Y_i(u^+), 0 \leq u \leq t\}$,
and $\bar{M}(t) = \sum_{i=1}^n M_i(t)$ is a martingale with respect to $\mathcal{F}_t = \cup_{i=1}^n \mathcal{F}_{t, i}$, the smallest $\sigma$-algebra containing all $\mathcal{F}_{t, i}$'s.

Our Cox's broken adaptive ridge (CoxBAR) estimation of $\bbeta$ starts with an initial Cox ridge regression estimator
\citep{verweij1994penalized}
\begin{equation}
\label{eq1:init_ridge}
\hat{\bbeta}^{(0)} = \mbox{arg}\min_{\bbeta} \left\{-2l_n(\bbeta) + \xi_n \sum_{j=1}^{p_n} \beta_j^2\right\},
\end{equation}
which is updated iteratively by a reweighed $L_2$-penalized Cox regression estimator
\begin{equation}
\label{eq1:l0_approx}
\hat{\bbeta}^{(k)}  = \mbox{arg} \min_{\bbeta}  \left\{-2l_n(\bbeta) + \lambda_n \sum_{j =1}^{p_n} \frac{\beta_j^2}{\left(\hat\beta_j^{(k-1)}\right)^2} \right\}, \quad k\ge 1.
\end{equation}
where  $\xi_n$ and $\lambda_n$ are non-negative penalization tuning parameters.
The CoxBAR estimator is defined as
\begin{equation}
\label{eq1:bar_lim}
\hat{\bbeta} = \lim_{k \to \infty} \hat{\bbeta}^{(k)}.
\end{equation}

Since $L_2$-penalization yields a non-sparse solution, defining the CoxBAR estimator as the limit is necessary to produce sparsity. Although $\lambda_n$ is fixed at each iteration, it is weighted inversely by the square of the ridge regression estimates from the previous iteration. Consequently, coefficients whose true values are zero will have larger penalties in the next iteration, whereas penalties for truly non-zero coefficients will converge to a constant.  We will show later in Theorem \ref{th1:oracle}  that under certain regularity conditions, the estimates of the truly zero coefficients shrink towards zero while the estimates of the truly non-zero coefficients converge to their oracle estimates.

\begin{remark}
\label{re:remark1} 
(Computation aspects of CoxBAR)
First of all, 
for moderate size data, one may calculate $\hat{\bbeta}^{(k)} $ in (\ref{eq1:l0_approx}) using the Newton-Raphson method as in \cite{frommlet2016adaptive} who outlined an iterative reweighted ridge regression for generalized linear models. 
 It appears at the first sight that (\ref{eq1:l0_approx}) will encounter numerical overflow as some of the coefficients 
$\hat\beta_j^{(k-1)}$ will go to zero as $k$ increases. However,  it can be shown that after some simple algebraic manipulations, the Newton-Raphson updating formula will only involve multiplications, instead of divisions, by $\hat\beta_j^{(k-1)}$s. So numerical overflow can be avoided. This further implies that once a $\hat\beta_j^{(k-1)}$ becomes zero, it will remain as zero in subsequent iterations. Thus one only needs to update $\hat{\bbeta}^{(k)} $ within the reduced nonzero parameter space, which is an appealing computational advantage for high dimensional settings.  Secondly, for massive size data with large $n$ and $p_n$, the Newton-Raphson procedure, which at each iteration calls for calculating both the gradient and Hessian,  can become practically infeasible  due to high computational costs, high memory requirements, and numerical instability.  In Section \ref{s1:comp} we will discuss how to adapt an  efficient algorithm for massive $L_2$-penalized Cox regression and exploit the sparsity in the covariate structure and the partial likelihood to make CoxBAR scalable to sHDMSS data.   
\end{remark}


\subsubsection{Oracle properties}
\label{s1:oracle}
We establish the oracle properties for the CoxBAR estimator for simultaneous variable selection and parameter estimation where we allow both $q_n$ and $p_n$ to diverge to infinity. 
Define
\begin{align*}
S^{(k)}(\bbeta, s) & = \frac{1}{n} \sum_{i=1}^n Y_i(s)\mathbf{x}_i(s)^{\otimes k} \exp \{\bbeta^T \mathbf{x}_i(s)\}, \hspace{.2in} k = 0, 1, 2, \\
\mathbf{E}(\bbeta, s) & = S^{(1)}(\bbeta, s) / S^{(0)}(\bbeta, s), \\
V(\bbeta, s) & = S^{(2)}(\bbeta, s)/ S^{(0)}(\bbeta, s) - \mathbf{E}(\bbeta, s)^{\otimes 2},
\end{align*}
where $Y_i(s) = I(\tilde{T}_i \geq s)$, $\mathbf{x}^{\otimes k} = 1, \mathbf{x}, \mathbf{x}\mathbf{x}^T$ for $k = 0, 1, 2$, respectively. 
Let $||\cdot||_p$ be the $L_p$-norm for vectors and the norm induced by the vector $p$-norm for matrices.  The following technical conditions will be needed in our derivations for the statistical properties of the CoxBAR estimator.
 \begin{description}
 \item[(C1)]  $\int_0^1 h_0(t) dt < \infty$;
 \item[(C2)]  There exists some compact neighborhood, $\mathcal{B}_0$, of the true value $\bbeta_0$ such that for $k = 0, 1, 2$, there exists a scalar, vector, and matrix function $s^{(k)}(\bbeta, t)$ defined on $\mathcal{B}_0 \times [0, 1]$ such that 
  \begin{align*}
\sup_{t \in [0,1], \bbeta \in \mathcal{B}_0} \left\| S^{(k)}(\bbeta, t)  - s^{(k)}(\bbeta, t) \right\|_2 = o_p(1), \quad \mbox{as $n\to \infty$};
\end{align*}
\item[(C3)] Let 
$s^{(1)}(\bbeta, t)  = \frac{\partial}{\partial \bbeta} s^{(0)}(\bbeta, t)$ and
$s^{(2)}(\bbeta, t) = \frac{\partial}{\partial \bbeta} s^{(1)}(\bbeta, t)$.
For $k = 0, 1, 2$, the functions $s^{(k)}(\bbeta, t)$ are continuous with respect to $\bbeta \in \mathcal{B}_0$, uniformly in $t \in [0, 1]$, and $s^{(k)}(\bbeta, t)$ are bounded; furthermore,  
$s^{(0)}(\bbeta, t)$ is bounded away from zero on $\mathcal{B}_0 \times [0, 1]$; \\

\item[(C4)] Let $e(\bbeta, t) = s^{(1)}(\bbeta, t) / s^{(0)}(\bbeta, t)$,
$v(\bbeta, t) = s^{(2)}(\bbeta, t)/s^{(0)}(\bbeta, t) - e(\bbeta, t)^{\otimes 2}$, and
$\Sigma(\bbeta) = \int_0^1 v(\bbeta, t)s^{(0)}(\bbeta, t)h_0(t)dt$.
There exists some constant $C_1 > 0$  such that
\begin{align*}
0 < C_1^{-1} < \eigen_{\min}\{\Sigma(\bbeta)\} \leq \eigen_{\max}\{\Sigma(\bbeta)\} < C_1 < \infty,
\end{align*}
uniformly in $\bbeta \in \mathcal{B}_0$, 
where for any matrix $A$, $\eigen_{\min}(A)$  and $\eigen_{\max}(A)$ represent its smallest and largest eigenvalues, respectively;
\item[(C5)] Let $\mathbf{U}_i = \int_0^1 \left\{ \mathbf{x}_i(t)-e(\bbeta_0, t) \right\} dM_i(t)$. There exists a constant $C_2$ such that $\sup_{1 \leq i \leq n} E(U_{ij}^2U_{il}^2) < C_2 < \infty$ for all $1 \leq j, l \leq p_n$, where $U_{ij}$ is the $j$-th element of $\mathbf{U}_i$;
\item[(C6)] 
As $n\to\infty$, $p_n^4/n \to 0$, $\lambda_n \to \infty$, $\xi_n \to \infty$, $\xi_nb_n/\sqrt{n}  \to 0,$ ${p_n/(na_n^2)} \to 0$,  $\lambda_nb_n^3\sqrt{q_n}/\sqrt{n} \to 0$ and $\lambda_n\sqrt{q_n}/(a_n^3\sqrt{n}) \to 0$, where $a_n=\min_{j=1, \ldots, q_n} (|\beta_{0j}|)$ and $b_n =\max_{j = 1, \ldots q_n} (|\beta_{0j}|) $. 
\end{description}
 Condition (C1) ensures a finite baseline cumulative hazard over the interval $[0, 1]$. Condition (C2) ensures the asymptotic stability of $S^{(k)}(\bbeta, t)$, as required for Cox regression under fixed dimension. Under diverging dimension, it follows from Theorem \ref{th1:oracle} of \cite{kosorok2007marginal} that under certain regularity conditions,
$ 
\sup_{t \in [0,1], \bbeta \in \mathcal{B}_0} \left\| S^{(k)}(\bbeta, t)  - s^{(k)}(\bbeta, t) \right\|_2 \leq \sqrt{p_n \ln p_n/n},
$
which implies that (C2) holds if $p_n \ln p_n /n \to 0$.
Condition (C3) is an asymptotic regularity condition similar to that for the fixed dimension Cox model.
 Condition (C4) guarantees that the covariance matrix of the score function is positive definite and has uniformly bounded eigenvalues for all $n$ and $\bbeta \in \mathcal{B}_0$. Other authors in the variable selection literature have also required a slightly weaker condition  \citep{fan2004nonconcave, cai2005variable, cho2013model, ni2016variable}. Condition (C5) is needed to prove the Lindeberg condition under diverging dimension in our proof. 
 Condition (C6) specifies the divergence or convergence rates for the  model size, the penalty tuning parameters, and the lower and upper bound of the true signal. 
 These technical assumptions are only sufficient conditions for our theoretical derivations and it is possible that our theoretical results hold
 under weaker conditions. For instance, we have observed in empirical studies that the CoxBAR method has good 
 performance even when $p_n$ is at the same order as $n$. Further efforts to relax these technical conditions are warranted in future research. 
 
\begin{theorem}[Oracle Properties]
\label{th1:oracle}
{\color{blue} 
Assume the regularity conditions (C1) - (C6) hold. 
Let  $\hat{\bbeta}_1$ and $\hat{\bbeta}_2$ be the first $q_n$ and the remaining $p_n-q_n$ components of  the CoxBAR estimator $\hat{\bbeta}$, respectively. 
Then, as $n\to\infty$,  
\begin{itemize}
\item[(a)] $P(\hat{\bbeta}_2=\boldsymbol{0})\to 1$;
\item[(b)] $\sqrt{n}\mathbf{b}_n^T\Sigma(\bbeta_0)_{11}^{-1/2}(\hat{\bbeta}_1 - \bbeta_{01}) \buildrel D \over \longrightarrow N( 0, 1)$, for any $q_n$-dimensional vector $\mathbf{b}_n$ such that $||\mathbf{b}_n||_2 \leq 1$ and where $\Sigma(\bbeta_0)_{11}$ is the first $q_n \times q_n$ submatrix of $\Sigma(\bbeta_0)$, where $\Sigma(\bbeta_0)$ is defined in Condition (C4).
\end{itemize}
}
\end{theorem}
\textcolor{blue}{Theorem \ref{th1:oracle}(a) establishes selection consistency of the CoxBAR estimator. Part (b) of the theorem
essentially states that the nonzero component of the CoxBAR estimator is asymptotically normal and equivalent to the weighted ridge estimator of the oracle model as shown in the proof provided in the Online Supplementary Material.}

\subsubsection{Grouping property}
When the true model has a group structure, it is desirable for a variable selection method to either retain or drop all variables that are clustered within the same group. Ridge regression has a grouping property for highly correlated covariates, and we show that the CoxBAR method has a similar grouping property since it is asymptotically equivalent to the weighted ridge estimator of the oracle model. 
 
 \begin{theorem}
\label{th1:group}
Assume that $X = (\mathbf{x}_i^T, \ldots \mathbf{x}_n^T)$ is standardized. That is, for all $j = 1, \ldots, p_n$,
$
\sum_{i=1}^n x_{ij} = 0,\  
\mathbf{x}_{[,j]}^T\mathbf{x}_{[,j]}  = n - 1,
$
where $\mathbf{x}_{[,j]}$ is the $j^{th}$ column of $X$. Suppose the regularity conditions (C1) - (C6) hold and let $\hat{\bbeta}$ be the CoxBAR estimator.
 Then for any $\hat{\beta}_i \neq 0$ and $\hat{\beta}_j \neq 0$,
\begin{equation}
\label{eq:group_bound}
|\hat{\beta}_i^{-1}-\hat{\beta}_j^{-1}|\leq \frac{1}{\lambda_n} \sqrt{2 \{ (n-1)(1 - r_{ij}) \}} \sqrt{n(1+d_n)^2} ,
\end{equation}
with probability tending to one, where $d_n = \sum_{i=1}^n \delta_i$, and  $r_{ij} = \frac{1}{n-1}\mathbf{x}_{[,i]}^T\mathbf{x}_{[,j]}$ is the sample correlation of $\mathbf{x}_{[,i]}$ and $\mathbf{x}_{[,j]}$.
\end{theorem}
We can see that as $r_{ij} \to 1$, the absolute difference between $\hat{\beta}_i$ and $\hat{\beta}_{j}$ approaches $0$, implying that the estimated coefficients of two highly correlated variables will be similar in magnitude.

\subsubsection{Selection of tuning parameters}
\label{s1:tuning}
A common strategy for tuning parameter selection in the penalized regression literature is to perform optimization
with respect to a data-driven selection criterion such as the $k$-fold cross-validation \citep{verweij1993cross},  Akaike information criterion (AIC) \citep{akaike1974new}, and Bayesian information criterion (BIC) \citep{schwarz1978estimating, volinsky2000bayesian, ni2018tuning}. While this strategy works for moderate sample size data, it is  
computationally costly  for massive sample size  data since multiple fits of the model are required. 
We point out that
the CoxBAR method has a distinct feature that it does not require costly data-driven search for an optimal pair of its tuning parameters, which is its key advantage in reducing the computational burden for fitting massive sample size survival data as illustrated later in Section 3.3 (Table 2). To this end, we first note that the objective function of
an $L_0$-penalized Cox regression with $\lambda_n=\ln(n)$ or $ \ln(d_n)\equiv \ln(\mbox{number of uncensored events})$ equals the BIC or censored BIC criterion, 
 respectively {\citep{schwarz1978estimating, volinsky2000bayesian, yang2005can}}. Hence the Cox-BAR estimator with a pre-specified $\lambda_n=\ln(n)$ or $ \ln(d_n)$ directly provides a local optima for  the BIC or censored BIC criterion, respectively.  
  We refer to the CoxBAR method with a prespecified $\lambda_n=\ln(n)$ or $ \ln(d_n)$ as BIC-CoxBAR or cBIC-CoxBAR, respectively, and illustrate in Section 3.2 (Table 2) that they have comparable or better performance as compared to some popular competing methods especially when the sample size is relatively large. 
 Secondly,  we demonstrate in Section \ref{s1:sens_sim} (Figure 1)
 that while fixing $\lambda_n$, the  BIC-CoxBAR and cBIC-CoxBAR estimators are insensitive to 
 $\xi_n$ over a wide interval (Figure 1). In practice, any small value of $\xi_n$ can used as long as 
the initial Cox ridge estimator can be numerically obtained.  

\subsection{Efficient implementation CoxBAR for sparse high-dimensional massive sample size (sHDMSS) data}
\label{s1:comp}
As mentioned earlier, the Newton-Raphson algorithm used for each iteration of the CoxBAR algorithm will become infeasible in large-scale settings with large $n$ and $p_n$ due to high computational costs, high memory requirements, and numerical instability.  
Because CoxBAR only involves fitting a reweighted Cox's ridge regression at each 
iteration step, it allows us to adapt an efficient algorithm developed by \cite{mittal2013high} for massive Cox ridge regression which among other techniques, include the column relaxation with logistic loss (CLG) algorithm using 1D updates with a one-step 
Newton-Raphson approximation and exploiting the sparsity in the covariate structure and the partial likelihood as detailed below.

\subsubsection{Adaptation of existing efficient algorithms for fitting massive $L_2$-penalized Cox's regression}
\label{s1:ccd}
\cite{mittal2013high} developed an efficient implementation of the massive  Cox's ridge regression for sHDMSS data. For parameter estimation, the authors adopted
the column relaxation with logistic loss (CLG) algorithm of \cite{zhang2001text}, which
 is a type of cyclic coordinate descent algorithm that estimates the coefficients using 1D updates. The CLG easily scales to high-dimensional data \citep{wu2008coordinate, simon2011regularization, gorst2012coordinate} and has been recently implemented for fitting
massive ridge and LASSO penalized generalized linear models \citep{suchard2013massive}, parametric survival models \citep{mittal2013large}, and Cox 's model \citep{mittal2013high}. 
When fitting this Cox ridge regression model, the CLG algorithm involves finding $\beta_j^{(new)}$, the value of the $j^{th}$ entry of $\bbeta$, that minimizes the negative penalized log-partial likelihood, $-l_p(\bbeta)$, assuming that the other values of $\beta_j$'s are held constant at their current values. For a Cox ridge regression with a penalty tuning parameters  $1/\phi_j $ for $j = 1, \ldots, p_n$,  finding $\beta_j^{(new)}$ is equivalent to finding the $z$ that minimizes,
\begin{align}
\nonumber
g(z) = -z \sum_{i = 1}^n\delta_i x_{ij} + \sum_{i = 1}^n \delta_i \ln \left\{ \sum_{y \in R(\tilde{T}_i)} \exp \left( \sum_{k = 1, k \neq j}^{p_n} \beta_k x_{yk} + z x_{yj} \right) \right\} + \frac{z^2}{2\phi_j},
\end{align}
where $R(\tilde{T}_i) = \{j: \tilde{T}_j > \tilde{T}_i\}$ is the risk set for observation $i$. Here we allow each parameter to be penalized differently. For example,   $\phi_j = (\hat\beta_j^{(k-1)})^2/\lambda_n$ in equation (\ref{eq1:l0_approx}) of the CoxBAR algorithm. Even for this 1D problem, an optimization procedure needs to be used since there is no closed form solution. Using a Taylor series approximation at the current $\beta_j$, one can approximate $g(\cdot)$ through
\begin{align}
\label{eq1:taylor}
g(z) \approx g(\beta_j) + g'(\beta_j)(z - \beta_j) + \frac{1}{2} g''(\beta_j)(z - \beta_j)^2,
\end{align}
where 
\begin{align}
\label{eq1:ccd_oned}
g'(\beta_j) = \left. \frac{d}{dz} g(z) \right\vert_{z = \beta_j} = -\sum_{i = 1}^n x_{ij} \delta_i + \sum_{i = 1}^n \delta_i \frac{\sum_{y \in R(\tilde{T}_i)} x_{yj}\exp(\bbeta^T\mathbf{x}_y)}{\sum_{y \in R(\tilde{T}_i)} \exp(\bbeta^T\mathbf{x}_y)} + \frac{\beta_j}{\phi_j},
\end{align}
and
\begin{align}
\label{eq1:ccd_twod}
g''(\beta_j) = \left. \frac{d^2}{dz^2} g(z) \right\vert_{z = \beta_j} & = \sum_{i = 1}^n \delta_i \frac{\sum_{y \in R(\tilde{T}_i)} x_{yj}^2 \exp(\bbeta^T\mathbf{x}_y)}{\sum_{y \in R(\tilde{T}_i)} \exp(\bbeta^T\mathbf{x}_y)} \\ \notag
& - \left( \sum_{i = 1}^n  \delta_i\frac{\sum_{y \in R(\tilde{T}_i)} x_{yj} \exp(\bbeta^T\mathbf{x}_y)}{\sum_{y \in R(\tilde{T}_i)} \exp(\bbeta^T\mathbf{x}_y)}  \right) + 
\frac{1}{\phi_j}.
\end{align}
Consequently,
the Taylor series approximation in Equation (\ref{eq1:taylor}) has its minimum at 
\begin{align}
\nonumber
\beta_j^{(new)} = \beta_j + \Delta \beta_j = \beta_j - \frac{g'(\beta_j)}{g''(\beta_j)}.
\end{align}
It is worth noting that as $\phi_j\to 0$, $g'(\beta_j)/g''(\beta_j) \to \beta_j$  and  thus $\beta_j^{(new)}\to 0$, which is an important feature of our CoxBAR algorithm as discussed in Remark \ref{re:remark1}. Furthermore, the above algorithm of \cite{mittal2013high} adopts multiple aspects of the work by \cite{zhang2001text} and \cite{genkin2007large}. For CLG, a trust region approach is implemented so that $|\Delta \beta_j|$  is not allowed to be too large on a single iteration. This prevents large updates in regions where a
quadratic is a poor approximation to the objective. Second, rather than iteratively updating $\beta_j^{(new)} = \beta_j + \Delta \beta_j$ until convergence, CLG does this only once before going on to the next variable. Since the optimal value of $\beta_j^{(new)}$ depends on the current value of the other $\beta_j$'s, there is little reason to tune each $\beta_j^{(new)}$ with high precision. Instead, we simply want to decrease $-l_p(\bbeta)$ before going on to the next $\beta_j$.

\subsubsection{Efficient computing and storage by accounting for sparsity in the covariate structure and partial likelihood }
Recall that the design matrix $X$ for sHDMSS data has few non-zero entries for each subject. Storing such a sparse matrix  as a dense matrix is inefficient and may increase computation time and/or cause a standard software to crash due to insufficient memory allocation. To the best of our knowledge, popular penalization packages such as {\textsc{glmnet}} \citep{friedman2010regularization} and {\textsc{ncvreg}} \citep{breheny2011coordinate} do not support a sparse data format as an input for right-censored survival models, although the former supports the input for other generalized linear models. For sHDMSS data, we propose to use specialized, column-data structures as in  \cite{suchard2013massive} and \cite{mittal2013high}. The advantage of this structure is two-fold: it significantly reduces the memory requirement needed to store the covariate information, and performance is enhanced  when employing cyclic coordinate descent. {For example when updating $\beta_j$, efficiency is gained when computing and storing the inner product $r_i = \bbeta^T\mathbf{x}_i$ using a low-rank update  $r_i^{(new)} = r_i + x_{ij} + \Delta \beta_j$ for all $i$ \citep{zhang2001text, genkin2007large, wu2008coordinate, suchard2013massive, mittal2013high}.}

Furthermore, as seen in equations (\ref{eq1:ccd_oned}) and (\ref{eq1:ccd_twod}) , one would need to calculate the series of cumulative sums introduced through the risk set $R(\tilde{T}_i) = \{j: \tilde{T}_j > \tilde{T}_i\}$ for each subject $i$. {These cumulative sums would need to be calculated when updating each parameter estimate in the optimization routine.} This can prove to be computationally costly, especially when both $n$ and $p_n$ are large.  By taking advantage of the sparsity of the design matrix, one can reduce the computational time needed to calculate these cumulative sums by entering into this operation only if at least one observation in the risk set has a non-zero covariate value along dimension $j$ and embarking on the scan at the first non-zero entry rather than from the beginning.  \cite{suchard2013massive} and \cite{mittal2013high} have implemented these efficiency techniques for conditional Poisson regression and Cox's regression, respectively.

Our CoxBAR implementation naturally exploits the sparsity in the data matrix and the partial likelihood by imbedding an adaptive version 
of \cite{mittal2013high}'s massive Cox's ridge regression  within each iteration of the iteratively reweighted Cox's ridge regression. We finally highlight that our CoxBAR method uses pre-specified tuning parameters as discussed in Section~\ref{s1:tuning}, which provides huge computation savings.

\subsection{CoxBAR for Ultrahigh-Dimensional Data}
\label{s1:ultrahigh}
The asymptotic properties of the CoxBAR estimator in the Section 2.1 are derived for $p_n<n$. In an ultrhigh dimensional setting where the number of covariates far exceeds the number of observations ($p_n>>n$), 
 one may couple a sure screening method with the CoxBAR estimator to obtain a two-step estimator with desirable selection and estimation properties.
There are a number of screening methods for right-censored survival data, which include marginal screening methods \citep{fan2010high, zhao2012principled,  gorst2013independent, song2014censored} and joint screening methods \citep{yang2016feature}.
For example, the sure independent screening (SIS) method of \citet{fan2010high} 
measures the importance of the covariates based on the marginal partial likelihood, which is fast, but may overlook important covariates that are jointly correlated, but not marginally correlated, with the observed survival time. 
The sure joint screening (SJS) method of \citet{yang2016feature} is based on the joint partial likelihood of potentially important covariates using a sparsity-restricted maximum partial likelihood estimate.  
Most of these methods have been shown to possess the sure screening property under certain regularity conditions in the sense that the subset of retained covariates includes the true model with probability tending to one.  

As an illustration, we consider a two-step estimator, referred to as SJS-CoxBAR, obtained by first performing the SJS method of \citet{yang2016feature} to reduce the covariate space to a subset $\hat s$ of  $m_n$ covariates and then fitting CoxBAR to the screened model $\hat s$. Specifically, let
 $\hat\bbeta = \sup_{\bbeta}\{l_n(\bbeta): ||\bbeta||_0 \le m_n \}$
be the sparsity-restricted maximum partial likelihood estimate of $\bbeta$ resulted from the iterative hard thresholding algorithm described in \citet{yang2016feature}. Define $\hat s=\{j: \hat\beta_{j} \neq 0 \}$. 
For simplicity, assume  that $\mathbf{x}$ is time independent. 
Below are additional conditions derived from \cite{yang2016feature} to ensure that $\hat{s}$ includes the 
 true model with sufficiently large probability. \\
 \begin{description}
 \item[(C7)] There exists $w_1, w_2 > 0$ and some non-negative constants $\tau_1$, $\tau_2$ such that $\tau_1 + \tau_2 < 1/2$ with $\min_{1 \leq j \leq q_n} |\beta_{0j}| \geq w_1n^{-\tau_1}$ and $q_n < m_n  \leq w_2 n^{\tau_2}$;
 \item[(C8)] There exists constants $c_1 > 0, \delta_1 > 0$ such that for sufficiently large $n$, $\mbox{eigen}_{\min}[H_n(\bbeta_0)] \geq c_1$ for $\bbeta_s \in \{\bbeta:||\bbeta_s - \bbeta_{0s}||_2 \leq \delta_1\}$ and $s \in S^{2m_n}_{+} \equiv \{s: s_0 \subset s;  ||s||_0 \leq 2m_n\}$, where $s_0 = \{j: \beta_{0j} \neq 0\}$; 
 \item[(C9)]  There exists $\delta_2 > 0$ such that $n^{-1/2} \sup_{i,t}   |\mathbf{x}_i|Y_i(t)I(\bbeta_0^T\mathbf{x}_i > \delta_2 |\mathbf{x}_i|) \stackrel{p}{\to} 0;$
 \item[(C10)] There exists constants $C_3, C_4 > 0$ such that $\max_{ij} |x_{ij}| < C_3$ and $\max_i |\mathbf{x}_i \bbeta_0| < C_4$.
 \item[(C11)] Let $t_1 < t_2 < \ldots, t_N$ be the ordered observed event times. There exists nonnegative constants $\gamma_j$ such that for every real number $t$,
 \begin{align*}
E\{\exp(tb_j)|\mathcal{F}_{t_{j-1}}\} \leq \exp(\gamma_j^2t^2/2),
\end{align*}
almost surely for $j = 1, 2, \ldots, N$. Further, for each $j$, define $\eta(b_j) = \min_j (\gamma_j)$. Now $|b_j| \leq K_j$ almost surely for $j = 1, \ldots, N$ and $E\{b_{j_1}, b_{j_2}, \ldots b_{j_k}\} = 0$ for $b_{j_1} < b_{j_2} < \ldots < b_{j_k}$, $k = 1, 2, \ldots$.
 \end{description}

\begin{theorem}
\label{th1:sjs}
Denote by $\hat{\bbeta}_{\hat s} =\left(\hat{\bbeta}_{\hat s 1}^T, \hat{\bbeta}_{\hat s 2}^T\right)^T$ the CoxBAR estimator of ${\bbeta}_{\hat{s}}$ obtained by fitting CoxBAR on the screened model $\hat{s}$, where
$\bbeta_s = \{\beta_j, j \in s\}$ for any subset $s$ of $\{1, \ldots, p_n\}$ and $\hat{\bbeta}_{\hat s 1}$ and 
$\hat{\bbeta}_{\hat s 2}$ represent the first $q_n$ and remaining $m_n-q_n$ components of $\hat{\bbeta}_{\hat s}$.
Suppose that conditions (C7) - (C11) hold and that conditions (C1) - (C6) hold for any submodel $s$ of size $m_n$.
In addition, assume that $\log p_n = O(n^{\kappa})$ for some $0 \leq \kappa < 1 - 2(\tau_1 + \tau_2)$.
Then 
\begin{itemize}
\item[(a)] (Sure screening property) $\Pr(s_0 \subset \hat{s}) \to 1$ as $n \to \infty$;
\item[(b)] (Oracle Property) Conditional on $s_0 \subset \hat{s}$, with probability tending to one,  $\hat{\bbeta}_{\hat s 2}=\boldsymbol{0}$,
and  $\sqrt{n}\mathbf{b}_n^T\Sigma(\bbeta_0)_{11}^{-1/2}(\hat{\bbeta}_1 - \bbeta_{01}) \buildrel D \over \longrightarrow N( 0, 1)$
for any $q_n$-dimensional vector $\mathbf{b}_n$ such that $||\mathbf{b}_n||_2 \leq 1$, and where $\Sigma(\bbeta_0)$ is defined in Condition (C4) with $p_n = m_n$.
\end{itemize}
\end{theorem}

\section{Simulations}
\label{s1:sim}
This section presents three simulation studies. 
First, we demonstrate in Section \ref{s1:sens_sim} that BIC-CoxBAR, the CoxBAR estimator with a fixed $\lambda_n= \ln(n)$, is insensitive to the tuning parameter $\xi_n$ of its initial ridge estimator and does well in terms of performing variable selection and correcting possible bias of the initial ridge estimator. 
Second, in Section \ref{s1:model_sim}, we evaluate and compare the operating characteristics of the BIC-CoxBAR estimator with some popular 
penalized Cox regression methods, where  we only consider settings with moderate sample sizes because most of the competing methods are inoperable for massive sample size data. 
Finally, in Section \ref{s1:large_sim}, we use a sHDMSS setting to illustrate the computational advantage of the BIC-CoxBAR estimator over its closest competitor. 


With the exception of Section \ref{s1:large_sim} we use the same simulation structure. 
Survival times are drawn from an exponential proportional hazards model  with baseline hazard $h_0(t) = 1$ and $\bbeta_0 = (0.20, 0, 0.35, 0, 0.50, 0.55, 0, 0, 0.70, 0.80, \mathbf{0}_{p_n - 10})$, representing small to moderate effect sizes. The design matrix $X = (\mathbf{x}_{1}^T, \ldots, \mathbf{x}_{n}^T)$ was generated from a $p_n$-dimensional normal distribution with mean zero and covariance matrix $\Sigma = (\sigma_{ij})$ with an autoregressive structure such that $\sigma_{ij} = 0.5^{|i - j|}$.  In Sections {\ref{s1:sens_sim}} and  {\ref{s1:model_sim}}, independent censoring times are simulated from a uniform distribution $U(0, u_{max})$, where $u_{max}$ is chosen to achieve $20\%$ censoring.

 \subsection{BIC-CoxBAR in action as $\xi_n$ varies}
\label{s1:sens_sim}

\begin{figure}[h]
\label{fig1:xi_plot}
\centering
\includegraphics[scale = 0.8]{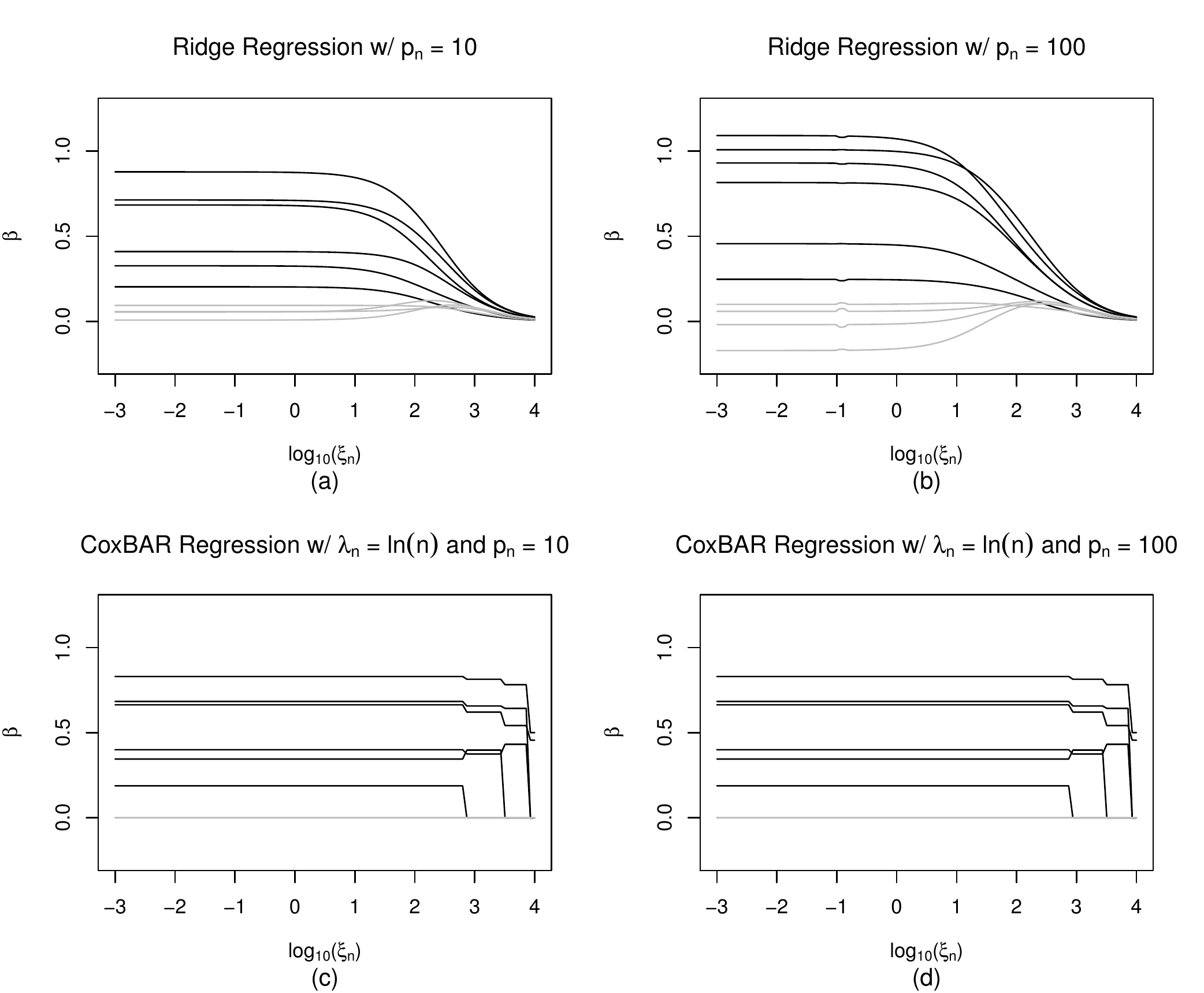}
\caption{
Path plot for CoxBAR regression with varying $\xi_n$ and $\lambda_n = \ln(n)$: (a) $p_n = 10$, (b) $p_n = 100$ for a random sample of size $n = 300$.
}

\end{figure}

While fixing $\lambda_n$ at  $\ln(n)$, as discussed in Section \ref{s1:tuning}, we illustrate below how the resulting BIC-CoxBAR estimator behaves by varying the tuning parameter
$\xi_n$ of the initial Cox ridge regression.   Figure \ref{fig1:xi_plot} (panels (c) and (d)) depicts the solution path plots of the BIC-CoxBAR estimator with respect to $\xi_n$ over a wide interval $[10^{-3}, 10^4]$ for $p_n=10$ and $p_n=100$ based on a random sample of size  $n = 300$. It is seen that over a large interval of $\xi_n$, the BIC-CoxBAR estimator is essentially unchanged, suggesting that there is no need to optimize over $\xi_n$ for a reasonable BIC-CoxBAR solution. Furthermore,  the BIC-CoxBAR estimator has correctly selected all nonzero coefficients and estimated all zero coefficients as zero; with essentially no estimation bias.

As a reference, we also display the  solution path plots of the corresponding initial ridge estimators in panels (a) and (b).
It is interesting to note that the initial ridge estimator starts to introduce over-shrinkage and consequently estimation bias 
when $\xi_n$ exceeds $10^1$. However, its bias has been effectively corrected by 
the BIC-CoxBAR until $\xi_n$ reaches a very large value of greater than $10^{2.8}$. The initial ridge estimator, especially for $p_n=100$, also displays large estimation bias for some of coefficients for all $\xi_n$, which has again been 
corrected by the BIC-CoxBAR estimator. Therefore, by iteratively refitting reweighted Cox ridge regression, the BIC-CoxBAR estimator not only performs variable selection
 by shrinking estimates of the true zero parameters to zero, but also effectively corrects the estimation bias from the initial Cox ridge estimator.
 
Similar results are obtained for cBIC-CoxBAR in our simulations which are not reported here.

\subsection{Model selection and parameter estimation}
\label{s1:model_sim}
In this simulation, we evaluate and compare the variable selection and parameter estimation performance of BIC-CoxBAR (CoxBAR with fixed $\lambda_n = \ln(n)$) and 
cBIC-CoxBAR (CoxBAR with fixed $\lambda_n = \ln(d_n)$  to CoxBAR(BIC), HARD(BIC) (hard-thresholding the Cox partial likelihood estimator), and three popular penalized Cox regression methods: LASSO(BIC)
\citep{tibshirani1997lasso}, SCAD(BIC) \citep{fan2002variable} , and adaptive LASSO (ALASSO(BIC)) \citep{zhang2007adaptive}, where BIC in parenthesis indicates that 
the BIC criterion was used to select the tuning parameters through a grid search.  We fix $\xi_n = 1$ for the CoxBAR methods since Section \ref{s1:sens_sim} suggests that the CoxBAR estimator is  insensitive to the selection of $\xi_n$. It is important to recognize the difference between BIC-CoxBAR and CoxBAR(BIC):  the former uses $\lambda_n = \ln(n)$, whereas the latter selects a tuning parameter $\lambda_n$ to minimize the BIC score.

Estimation bias is summarized through the sum of squared bias (SSB), $E\{\sum_{i=1}^{p_n} (\hat{\bbeta}_i - \bbeta_{0i})^2\}$. Variable selection performance is measured by a number of indices: the mean number of false positives (FP), the mean number of false negatives (FN); probability that the selected model is equal to the true model (TM); AIC value, BIC value, and the average number of variables that are correctly ranked (ACR). We also include the inclusion probability for each of the nonzero coefficients. All simulations were conducted using R. Hard thresholding was performed using the {\textsc{coxph}} function in the {\textsc{survival}} package.  We use the R packages {\textsc{glmnet}} for LASSO and adaptive LASSO (ALASSO), and {\textsc{ncvreg}} for SCAD in our simulations. For ALASSO, we let the initial estimator be the maximum partial likelihood estimator since $p_n < n$.  Part of the simulation results are summarized in Table 1
where we fix $n = 300, 1000$ and $p_n = 100$. For each scenario, 100 replications are conducted.
We actually considered a variety of combinations of $n$ and $p_n$ and as well as different data-driven tuning parameter selection criteria such as cross-validation \citep{verweij1993cross} and GIC \citep{ni2018tuning}. The results are consistent with Table 1 and thus not included here.

\begin{table}
\caption{(Moderate dimension and sample size) Simulated estimation and variable selection performance of BIC-CoxBAR (CoxBAR with $\lambda_n = \ln(n)$) and
cBIC-CoxBAR (CoxBAR with $\lambda_n = \ln(d_n)$), along with CoxBAR(BIC), HARD(BIC), LASSO(BIC), SCAD(BIC), and ALASSO(BIC) where
BIC in parenthesis indicates that 
the BIC criterion was used to select the tuning parameters via a grid search. 
(SSB = sum squared bias; 
$P_j$ = probability that $\beta_{0j}$ is correctly identified;
FN = mean number of false positives; FP = mean number of false negatives; 
TM = probability that the selected model is equal to the true model; AIC = AIC score;
BIC = BIC score;
ACR = average number of correctly ranked non-zero covariates; 
Each entry is based on 100 Monte Carlo samples of size $n=300, 1000$, $p_n = 100$, censoring rate = $20\%$.) } \label{tab1:s32_results}
\centering
\setlength{\tabcolsep}{2.5pt}
\fbox{%
\begin{tabular}{c|ccccccccccccc}
$n = 300$ & SSB & $P_1$ & $P_3$ & $P_5$ & $P_6$ & $P_9$ &$P_{10}$ & FN & FP & TM & AIC & BIC & ACR \\ 
  \hline
BIC-CoxBAR & {\bf 0.09} & 0.27 & 0.92 & 1.00 & 1.00 & 1.00 & 1.00 & 0.81 & {\bf 0.09} & \bf 0.22 & 2055.42 & 2074.98 & 3.83 \\ 
  cBIC-CoxBAR & \bf 0.09 & 0.29 & 0.94 & 1.00 & 1.00 & 1.00 & 1.00 & 0.77 & \bf 0.11 & \bf 0.25 & 2054.83 & 2074.61 & 3.83 \\ 
  \hline
  CoxBAR(BIC) & 0.11 & 0.59 & 0.99 & 1.00 & 1.00 & 1.00 & 1.00 & 0.42 & 1.59 & 0.15 & 2043.64 & 2070.20 & 4.04 \\ 
  HARD(BIC) & 0.64 & 0.19 & 0.74 & 0.95 & 0.98 & 1.00 & 1.00 & 1.14 & 1.11 & 0.05 & 2105.05 & 2127.16 & 2.97 \\ 
  LASSO(BIC) & 0.22 & 0.82 & 1.00 & 1.00 & 1.00 & 1.00 & 1.00 & 0.18 & 3.15 & 0.02 & 2081.52 & 2114.75 & 3.97 \\ 
  SCAD(BIC) & 0.14 & 0.75 & 1.00 & 1.00 & 1.00 & 1.00 & 1.00 & 0.25 & 2.17 & 0.11 & 2059.99 & 2089.32 & 3.47 \\ 
  ALASSO(BIC) & 0.12 & 0.49 & 0.97 & 1.00 & 1.00 & 1.00 & 1.00 & 0.54 & 1.77 & 0.09 & 2059.15 & 2085.93 & 3.84 \\  
  \hline
  \hline
    $n = 1000$ & & & & & & & & & & & & & \\
\hline
  BIC-CoxBAR & \bf 0.02 & 0.93 & 1.00 & 1.00 & 1.00 & 1.00 & 1.00 & 0.07 & \bf 0.00 & \bf 0.93 & 8731.86 & 8760.96 & 5.04 \\ 
  cBIC-CoxBAR & \bf 0.02 & 0.93 & 1.00 & 1.00 & 1.00 & 1.00 & 1.00 & 0.07 & \bf 0.01 & \bf 0.93 & 8731.69 & 8760.84 & 5.04 \\ 
  \hline
  CoxBAR(BIC) & 0.02 & 0.98 & 1.00 & 1.00 & 1.00 & 1.00 & 1.00 & 0.02 & 0.72 & 0.55 & 8725.74 & 8758.63 & 5.08 \\ 
  HARD(BIC) & 0.04 & 0.93 & 1.00 & 1.00 & 1.00 & 1.00 & 1.00 & 0.07 & 0.33 & 0.75 & 8737.61 & 8768.33 & 5.00 \\ 
  LASSO(BIC) & 0.08 & 1.00 & 1.00 & 1.00 & 1.00 & 1.00 & 1.00 & 0.00 & 2.90 & 0.21 & 8768.42 & 8812.09 & 5.02 \\ 
  SCAD(BIC) & 0.02 & 0.98 & 1.00 & 1.00 & 1.00 & 1.00 & 1.00 & 0.02 & 0.48 & 0.60 & 8736.51 & 8768.21 & 4.94 \\ 
  ALASSO(BIC) & 0.02 & 0.98 & 1.00 & 1.00 & 1.00 & 1.00 & 1.00 & 0.02 & 0.40 & 0.70 & 8734.18 & 8765.49 & 5.02 \\ 
  \hline
  \end{tabular}
}
\end{table}

It is observed from Table 1 
that when the tuning parameter $\lambda$ is selected by minimizing the BIC score as the other methods, the performance of CoxBAR(BIC) is generally comparable to other methods with respect to all measures across all scenarios. We further examine the performance of BIC-CoxBAR, the CoxBAR method with a fixed $\lambda_n= \ln(n)$. For the smaller sample size $n=300$,  while exhibiting similar performance to other methods with respect to most measures, the BIC-CoxBAR estimator shows a lower number of false nonzeros (FP), lower estimation bias (SSB), slightly lower probability ($P_1$) of retaining the weak signal $\beta_1$, and a substantially higher probability of selecting the exact true model (TM).  For the larger sample size $n=1000$, BIC-CoxBAR with a fixed $\lambda_n = \ln(n)$ performs equally well as other methods with respect to all measures except that it remains to show a much higher probability of selecting the exact true model (TM).  This makes BIC-CoxBAR a better choice for fitting large-scale sHDMSS data since in addition to comparable or better performance, it does not require costly data-driven tuning parameter selection and thus has an computational advantage as shown later in Section 3.3.

We also investigated the performance of the two-stage SJS-CoxBAR estimator described in 
Section \ref{s1:ultrahigh} in ultrahigh dimensional settings where $p_n$ is much larger than $n$. The results are given in  Online Supplementary Material \ref{ap1:hd_sim} with similar messages except that  the methods using data-driven tuning parameter selection have an overwhelming number of false positives which, as a consequence, inflates the estimation bias.  

\subsection{Sparse high-dimensional massive sample size data}
\label{s1:large_sim}
In this simulation, we simulate a sHDMSS dataset with $n = 200000$ and $p_n = 20000$. Survival times are generated from an exponential hazards model with baseline hazard $h_0(t) = 1$ and regression coefficients $\bbeta_0 = (\mathbf{0.7}_{10}, \mathbf{0.5}_{10}, \mathbf{1}_{10}, \mathbf{-0.7}_{10}, \mathbf{-0.5}_{10},  \mathbf{-1}_{10}, \mathbf{0}_{p_n - 60})$. We set the censoring rate to $95\%$ and the covariates sparseness level to  $98\%$ such that each row of $X$ has, on average, only $2\%$ of the entries being assigned a non-zero value. The estimated amount of memory being used to store this dense design matrix is over 16GB, which exceeds the functional capacity of most statistical software packages and standard hardware. To overcome this difficulty, we efficiently store the information in a coordinate list fashion which only requires approximately 1GB of memory. We compare our CoxBAR method with the massive sparse Cox's regression for LASSO  (mCox-LASSO) using the {\textsc{Cyclops}} package \citep{suchard2013massive, mittal2013high} which, to the best of our knowledge, is the fastest software available today that exploits the sparsity of the large-scale survival data for efficient computing 
and
offers $>$ 10-fold speedup \citep{mittal2013high} over its competitors such as \textsc{CoxNet} \citep{simon2011regularization} and \textsc{FastCox} \citep{yang2012cocktail}. 
For LASSO, cross validation (mCox-LASSO (CV)), combined with
a nonconvex optimization technique which is more efficient than the classical grid search approach, and BIC score minimization (mCox-LASSO (BIC)), implemented with the classical grid search approach, were used to find the optimal value for the tuning parameter. For the CoxBAR method, we considered $\lambda_n = \ln(n)$ (BIC-CoxBAR) and 
$\lambda_n =\ln(d_n)$ (cBIC-CoxBAR) while fixing $\xi_n = 1$. The results are summarized in Table 2.
\begin{center}
\begin{table}
\caption{(High dimensional and massive sample size) Runtime, estimation, and variable selection results of BIC-CoxBAR (CoxBAR with $\lambda_n = \ln(n)$), 
cBIC-CoxBAR (CoxBAR with $\lambda_n = \ln(d_n)$), and the massive Cox regression with LASSO penalty (mCox-LASSO,
 \cite{mittal2013high}) for a simulated sHDMSS dataset with $n = 200,000$ and $p_n = 20,000$.
(SSB = sum squared bias;
FP= number of false positives; FN = number of false negatives; BIC = BIC Score.) } 
\label{tab1:s33_results}
\centering
\fbox{%
\begin{tabular}{l | r  r r r r}
 Method & Runtime (minutes)  & SSB & FP & FN &  BIC  \\ 
  \hline
BIC-CoxBAR & {{32}} & 1.17 & 0 & 2 & 226262.8 \\ 
cBIC-CoxBAR & 33 & {{0.65}} & 1 & 0 & {226217.2} \\ 
mCox-LASSO (CV) & 148 & 4.12 & 120 & 0 & 227955.3 \\ 
mCox-LASSO (BIC) & 164 & 6.18 & 5 & 0 & 227059.5 
\end{tabular}
}
\end{table}
\end{center}
We observed that both mCox-LASSO methods have retained all 60 true nonzero coefficients together with a moderate to large number of noise variables (5 for BIC and 120 for CV). In contrast, BIC-CoxBAR 
selected all but two of the weakest signals with no noise variables and cBIC-CoxBAR retains all 60 nonzero coefficients with only 1 noise variable. 
As expected, both BIC-CoxBAR and cBIC-CoxBAR have much smaller parameter estimation bias (SSB $\approx 1.17$ and SSB $\approx 0.65$, respectively) than mCox-LASSO (SSB $\approx 4.12$ for CV and SSB $\approx 6.18$ for BIC). 
Moreover,  although optimized in the {\textsc{Cyclops}} package, mCox-LASSO took at least 148 minutes to run while  BIC-CoxBAR or  cBIC-CoxBAR only took around 32 minutes, which represents a five-fold speedup. Finally, for model fit,both CoxBAR methods have much lower BIC scores compared to the mCox-LASSO methods. In summary, this simulation confirms that the CoxBAR methods are superior to mCox-LASSO in terms of obtaining a more sparse and accurate model, reducing estimation bias, offering better model fit with smaller BIC scores, and most importantly, reducing the computation time substantially with about 5-fold speedup. 

We further examined the solution paths of mCox-LASSO and CoxBAR in Figure 2, where 
the solid and dashed lines in the mCox-LASSO solution path plot (Figure 2(a)) represent the estimates at the optimal tuning parameter obtained via cross validation and BIC minimization, respectively. We can see that the mCox-LASSO solution path changes rapidly as its tuning parameter varies. Thus it is important to use an optimal data-driven selected tuning parameter for mCox-LASSO, which is computationally intensive for sHDMSS data. Also, mCox-LASSO tends to keep a substantial number of noise variables with large estimation bias even at its optimal penalty value using various criteria. In contrast, the CoxBAR solution path plot (Figure 2(b)) with respect to $\lambda_n$ changes very slowly over a relative large interval that includes $\ln(n)$ (black solid vertical line) and $\ln(d_n)$ (black dotted vertical line),  and correctly selects the true model with small estimation bias. 
For the CoxBAR method, we also made a CoxBAR solution path plot with respect to $\xi_n$, while fixing $\lambda_n = \ln (n)$ 
in Figure 2(c). It shows that the CoxBAR estimates are very stable and, in fact, almost correctly identify the exact true model over a large range of 
$\xi_n$, affirming our observation in Section \ref{s1:sens_sim} with small scale data. 
\begin{figure}[h]
\centering
\label{fig1:largescale_plot}
\includegraphics[scale = 0.6]{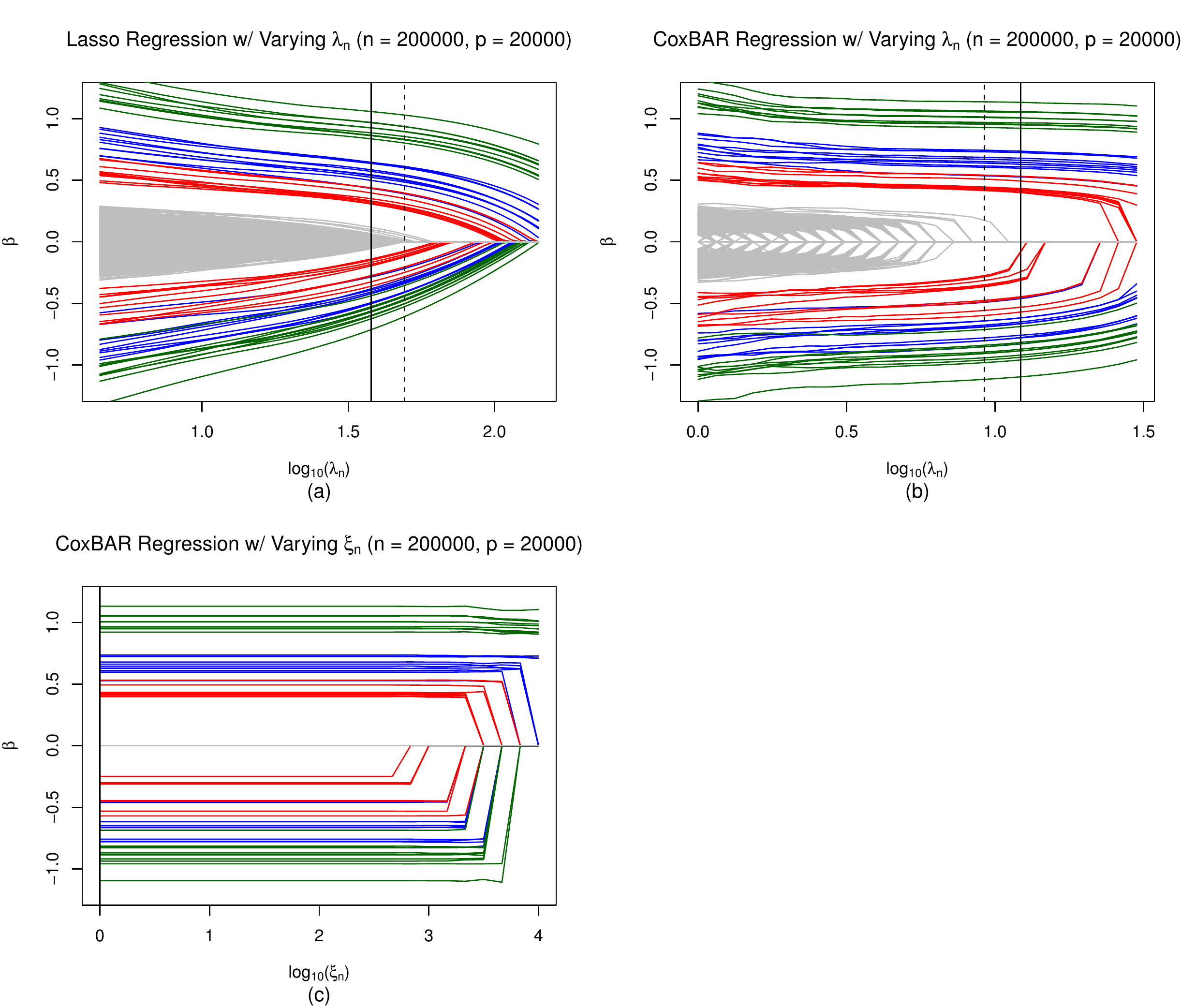}
\caption{
Path plots for mCox-LASSO and CoxBAR regression: (a) Path plot for mCox-LASSO regression, where the black dashed line represents the estimates when using cross validation to find the optimal value of the tuning parameter;   (b) Path plot for CoxBAR regression with $\xi_n = 1$ and varying $\lambda_n$, where the black solid and dashed line represent estimates for $\lambda_n = \ln(n)$ and $\lambda_n = \ln(d_n)$, respectively; 
(c) Path plot for CoxBAR regression with $\lambda_n = \ln(n)$ and varying $\xi_n$, where  the black solid line represent the estimates for CoxBAR when $\xi_n = 1$. }
\end{figure}

\section{A real data example}
\label{s1:data}


For an application of CoxBAR regression in the large-scale sparse data setting, we consider a subset of the National Trauma Data Bank that involves children and adolescents. This dataset was previously analyzed by  \cite{mittal2013high} as an example for efficient massive Cox regression with LASSO (mCox-LASSO) and ridge regression to sparse high-dimensional and massive sample size (sHDMSS) data. The dataset includes 210,555 patient records of injured children under 15 that were collected over 5 years (2006 -2010). Each patient record includes 125,952 binary covariates which indicate the presence, or absence, of an attribute (ICD9 Codes, AIS codes, etc.) as well as the two-way interactions between attributes. The outcome of interest is mortality after time of injury. The data is extremely sparse, with less than $1\%$ of the covariates being non-zero and has a censoring rate of $98\%$. Since the data is too large to fit other popular oracle procedures, {we compare the CoxBAR method, with $\lambda_n = \ln(n)$ and $\lambda_n = \ln(d_n)$ and with $\xi_n = 1$,} to mCox-LASSO with cross validation and BIC score minimization. We run both models on the full dataset and record the partial log-partial likelihood, number of non-zero covariates, BIC score, and computing time in Table \ref{tab1:ntdb}. 

As shown in Table \ref{tab1:ntdb}, the BIC-CoxBAR and cBIC-CoxBAR methods select far fewer covariates than mCox-LASSO with a three to six-fold speedup in computing time. Both CoxBAR methods took less than a day to run while mCox-LASSO took about three to five days to finish.  Of the 120 covariates selected by mCox-LASSO (BIC), BIC-CoxBAR and cBIC-CoxBAR also selected 48 and 60 of those, respectively. The covariates selected by BIC-CoxBAR are also a subset of the covariates identified by cBIC-CoxBAR. Further, the BIC score for the two CoxBAR methods are substantially smaller than those of the mCox-LASSO methods. In summary, the BIC-CoxBAR and cBIC-CoxBAR methods  identify fewer non-zero covariates with a significant reduction in computation time and improvement in model selection performance.

\begin{table}
\caption{(Pediatric NTDB data)  Comparison of mCox-LASSO 
and CoxBAR regression
for the pediatric NTDB data. (mCox-LASSO (CV) and mCox-LASSO (BIC) correspond to mCox-LASSO using cross validation and BIC selection criterion, respectively. BIC-CoxBAR and cBIC-CoxBAR denote CoxBAR with $\lambda_n = \ln(n)$ and $\lambda_n = \ln(d_n)$ respectively)}
\label{tab1:ntdb}
\centering
\fbox{%
\begin{tabular}{r | r r r r}
 Method & Runtime (in hours) & Log-partial likelihood & $\#$ Selected & BIC Score \\ 
  \hline
mCox-LASSO (CV) & 76 & -32682.17 & 186 & 67644.23 \\ 
mCox-LASSO (BIC) & 115 & -32969.44 & 
120  & 67368.56 \\
BIC-CoxBAR & 18 & -32814.54 & 55 & 65897.66 \\ 
cBIC-CoxBAR & 21  & -32517.85 & 81 & 66028.56 \\ 
\end{tabular}
}
\end{table}

\section{Discussion}
\label{s1:disc}

Although there are available many penalized Cox regression methods for simultaneous variable selection and parameter estimation, most current algorithms and softeware will grind to a halt and become inoperable for sHDMSS data. 
We have developed a new  sparse Cox regression method, named CoxBAR, 
by iteratively performing reweighted $L_2$-penalized Cox regression where the penalty is adaptively reweighted to approximate the $L_0$-penalty.  The resulting CoxBAR estimator can be viewed as a special local $L_0$-penalized Cox regression method and is shown to enjoy the best of $L_0$- and $L_2$-penalized Cox regression: it is selection consistent, oracle for parameter estimation, stable, and scalable to high-dimensional covariates, and has a grouping property for highly correlated covariates. We illustrate through empirical studies that the CoxBAR estimator has  comparable or better performance for variable selection and parameter estimation as compared to current penalized Cox regression methods, and most importantly, it has a distinct computational advantage with a 5-fold speedup over its closest competitor for sHDMSS data. Its computing efficiency is primarily due to the facts that the CoxBAR algorithm allows us to easily adapt existing efficient algorithms and software for massive $L_2$-penalized Cox regression \citep{mittal2013high}  and that it does not require costly data-driven tuning parameter selection. 

In addition to its application to sHDMSS data, our developed theory for CoxBAR guarantees that it can be combined with a sure screening procedure to obtain a conditional oracle  sparse regression method for  ultrahigh dimensional data when the dimension  far exceeds the sample size. 
It is also worth noting that
our $L_0$-based CoxBAR method and theory can be easily extended to an $L_d$-based CoxBAR method for any $d \in [0, 1]$, by replacing $(\hat{\beta}_j^{(k-1)})^2$ with $|\hat{\beta}_j^{(k-1)} |^{2 - d}$ in (\ref{eq1:l0_approx}).  We have observed empirically that
as $d$ increases towards 1,  the resulting estimator becomes less sparse, and the average number of false positives as well as estimation bias tend to increase, especially for larger 
$p_n$, while the average number of false negatives tends to decrease.  In practice,
$d$ can be used as a resolution tuning parameter. Finally, the proposed CoxBAR method can extended to obtain scalable sparse regression methods for more complex sampling schemes such as cohort sampling, which is currently under investigation by our team.

\vskip 10pt
\noindent {\large\bf Acknowledgement}
\vskip 5pt
The authors are grateful to Professor Piotr Fryzlewicz, the associate editor, and the referees for their insightful comments and suggestions that have greatly improved the paper. Gang Li's research was supported in part by National Institute of Health Grants P30 CA-
16042, UL1TR000124-02, and P01AT003960.

\newpage

\bibliographystyle{chicago}
\bibliography{coxbar_jrssb}

\newpage
\appendix
\pagenumbering{arabic}

\numberwithin{equation}{section}
\numberwithin{figure}{section}
\numberwithin{table}{section}



\section{Online Supplementary Material}

\subsection{Proof of Theorem \ref{th1:oracle}}
\label{pr1:oracle}

To prove Theorem \ref{th1:oracle}, we first establish five lemmas. 

\begin{lemma}[Asymptotic Variance of $\mathbf{U}_i$]
\label{lem1:asymvar}
 Let $\mathbf{U}_i = \int_0^1 \left\{ \mathbf{x}_i(t)-e(\bbeta_0, t) \right\} dM_i(t)$ be defined as in Condition (C5) and $\Sigma = \Sigma(\bbeta_0) =  \int_0^1 v(\bbeta_0, t)s^{(0)}(\bbeta_0, t)h_0(t)dt$, $\mathbf{e}(\bbeta_0, t)$,  and $v(\bbeta_0, t)$ be defined as in Condition (C4). Then under Conditions (C1) - (C4), 
 \begin{align}
\label{eq1:u_conv}
\left\| \frac{1}{n} \sum_{i=1}^n Var(\mathbf{U}_i) - \Sigma \right\|_2 = o_p(1),
\end{align}
as $n \to \infty$.
\end{lemma}

{\bf{Proof. }}  Denote by $U_{ij}$  the $j^{th}$ element of $\mathbf{U}_i$ and $e_j(\bbeta_0, s)$ as the $j^{th}$ element of $\mathbf{e}(\bbeta_0, s)$. Then,
\begin{align*}
Cov(U_{ij}, U_{ik}) & = \left\langle \int_0^1\{x_{ij}(s) - e_j(\bbeta_0, s)\}dM_i(s), \int_0^1 \{x_{ik}(s) - e_k(\bbeta_0, s)\}dM_i(s) \right\rangle \\
& = \int_0^1\{x_{ij}(s) - e_j(\bbeta_0, s)\}\{x_{ik}(s)-e_k(\bbeta_0, s)\}Y_i(s)h_0(s)\exp\{\bbeta^T \mathbf{x}_i(s)\}ds.
\end{align*}
Hence,
 \begin{align*}
\frac{1}{n} \sum_{i=1}^n Var(\mathbf{U}_i) & = \int_0^1  \frac{1}{n}  \sum_{i=1}^n h_0(s) Y_i(s) \mathbf{x}_i(s)^{\otimes 2} \exp\{\bbeta_0^T \mathbf{x}_i(s)\}ds \\
& - \int_0^1\frac{1}{n} \sum_{i=1}^n h_0(s) Y_i(s) \mathbf{x}_i(s) \mathbf{e}(\bbeta_0, s)^T \exp(\bbeta_0^T\mathbf{x}_i(s)\} ds \\
& - \int_0^1 \frac{1}{n} \sum_{i=1}^n h_0(s) Y_i(s) \mathbf{e}(\bbeta_0, s)  \mathbf{x}^T_i(s)\exp\{\bbeta_0^T\mathbf{x}_i(s)\} ds \\
& + \int_0^1   \mathbf{e}(\bbeta_0, s)^{\otimes 2}   \frac{1}{n} \sum_{i=1}^n h_0(s) Y_i(s) \exp(\bbeta_0^T\mathbf{x}_i(s)\} ds \\
& = \int_0^1 S^{(2)}(\bbeta_0, s)h_0(s)ds - \int_0^1 S^{(1)}(\bbeta_0, s) \mathbf{e}(\bbeta_0, s)^T h_0(s)ds \\
& - \int_0^1 \mathbf{e}(\bbeta_0, s) S^{(1)}(\bbeta_0, s)^T h_0(s)ds + \int_0^1 \mathbf{e}(\bbeta_0, s)^{\otimes 2}S^{(0)}(\bbeta_0, s) h_0(s)ds.
\end{align*}
Also note that 
\begin{align*}
 \Sigma(\bbeta_0) & = \int_0^1 v(\bbeta_0, s) s^{(0)}(\bbeta_0, s)h_0(s)ds \\
 & =  \int_0^1 \left\{  \frac{s^{(2)}(\bbeta_0, s)}{s^{(0)}(\bbeta_0, s)} -  \mathbf{e}(\bbeta_0, s)^{\otimes 2}\right\} {s^{(0)}(\bbeta_0, s)}h_0(s)ds \\
& =  \int_0^1 s^{(2)}(\bbeta_0, s)h_0(s)ds - \int_0^1 s^{(1)}(\bbeta_0, s) \mathbf{e}(\bbeta_0, s)^T h_0(s)ds \\
& - \int_0^1 \mathbf{e}(\bbeta_0, s) s^{(1)}(\bbeta_0, s)^T h_0(s)ds + \int_0^1 \mathbf{e}(\bbeta_0, s)^{\otimes 2}s^{(0)}(\bbeta_0, s) h_0(s)ds,
\end{align*}
since $e(\bbeta_0, t) = s^{(1)}(\bbeta_0, t) / s^{(0)}(\bbeta_0, t)$. Therefore, \begin{align*}
\left\| \frac{1}{n} \sum_{i=1}^n Var(\mathbf{U}_i)  - \Sigma(\bbeta_0) \right\|_2 & \leq 
\left\|  \int_0^1 \left\{S^{(2)}(\bbeta_0, s) - s^{(2)}(\bbeta_0, s) \right\} h_0(s)ds \right\|_2 \\
& + \left\| \int_0^1 \left\{S^{(1)}(\bbeta_0, s) - s^{(1)}(\bbeta_0, s) \right\} \mathbf{e}(\bbeta_0, s)^T h_0(s)ds \right\|_2 \\
& + \left\| \int_0^1 \mathbf{e}(\bbeta_0, s) \left\{S^{(1)}(\bbeta_0, s) - s^{(1)}(\bbeta_0, s) \right\}^T h_0(s)ds \right\|_2 \\
& + \left\|  \int_0^1 \mathbf{e}(\bbeta_0, s)^{\otimes 2} \left\{S^{(0)}(\bbeta_0, s) - s^{(0)}(\bbeta_0, s) \right\} h_0(s)ds \right\|_2 \\
& = o(1),
\end{align*}
where the last step follows from Conditions (C1), (C2), and (C3).$\qquad\Box$

\begin{lemma}[Asymptotic Normality of the Score Function]
\label{lem1:consistency}
Let $l_n(\bbeta)$ be the log-partial likelihood as defined in (\ref{eq1:loglik}). For any $p_n$-dimensional vector $\mathbf{d}_n$ such that $||\mathbf{d}_n||_2 = 1$, under Conditions (C1) - (C6), we have
\begin{align}
\label{eq1:mle}
n^{-1/2} \mathbf{d}_n^T\Sigma(\bbeta_0)^{-1/2} \dot{l}_{n}(\bbeta_0) \stackrel{D}{\rightarrow}  N(0, 1),
\end{align}
where $\dot{l}_n(\bbeta_0)$ is the first derivative of $l_n(\bbeta_0)$ and $\Sigma(\bbeta_0)$ is defined in Condition (C4). 
\end{lemma}

{\bf{Proof: }} First, observe that 
\begin{align}
\dot{l}_{n}(\bbeta_0) 
& = \sum_{i=1}^n \int_0^1 \left\{ \mathbf{x}_i(t) - \mathbf{E}(\bbeta_0, s) \right\} dM_i(s) \nonumber\\
& = \sum_{i=1}^n \int_0^1 \left\{ \mathbf{x}_i(t) - \mathbf{e}(\bbeta_0, s) \right\} dM_i(s) - \sum_{i=1}^n \int_0^1 \left\{ \mathbf{E}(\bbeta_0, s) - \mathbf{e}(\bbeta_0, s) \right\} dM_i(s) \nonumber\\
& = \sum_{i=1}^n \mathbf{U}_i + o_p(\sqrt{n}), \label{U1}
\end{align}
where $\mathbf{U}_{i}$ is defined as in condition (C4), and the right-hand side of the last equality is due to $||\mathbf{E}(\bbeta_0, s) - \mathbf{e}(\bbeta_0, s)||_2 \to o_p(1)$ from conditions (C2) and (C3), and $n^{-1/2} \sum_{i=1}^n \int_0^1 dM_i(s) = O_p(1)$.
Therefore
\begin{align*}
n^{-1/2} \mathbf{d}_n^T \Sigma(\bbeta_0)^{-1/2} \dot{l}_{n}(\bbeta_0) = \sum_{i=1}^n Y_{ni} + o_p(1),
\end{align*}
 where $Y_{ni} = n^{-1/2} \mathbf{d}_n^T \Sigma(\bbeta_0)^{-1/2} \mathbf{U}_{i}$. Note that $Y_{ni}$ has mean zero and
 \begin{align*}
s_n^2 = \sum_{i=1}^n Var(Y_{ni}) & = \frac{1}{n} \sum_{i=1}^n  \mathbf{d}_n^T \Sigma(\bbeta_0)^{-1/2} Var(\mathbf{U}
_{i})\Sigma(\bbeta_0)^{-1/2}\mathbf{d}_n  \\
& = \mathbf{d}_n^T \Sigma(\bbeta_0)^{-1/2} \left\{ \frac{1}{n} \sum_{i=1}^n  Var(\mathbf{U}
_{i}) \right\} \Sigma(\bbeta_0)^{-1/2}\mathbf{d}_n \to 1,
\end{align*}
where the last step follows from Lemma \ref{lem1:asymvar}. Hence by the Lindeberg-Feller central limit theorem, 
 \begin{align}
\label{eq1:lindeberg_clt}
\frac{\sum_{i=1}^n Y_{ni}}{s_n} \stackrel{D}{\rightarrow} N(0, 1),
\end{align}
if the following Lindeberg condition for $Y_{ni}$ holds: for all $\epsilon > 0$,
\begin{align}
\label{eq1:lindeberg_cond}
\frac{1}{s_n^2} \sum_{i=1}^n E\{Y_{ni}^2 I(|Y_{ni}| \geq \epsilon s_n)\}  \to 0,
\end{align}
as $n \to \infty$. To verify (\ref{eq1:lindeberg_cond}) we note that

\begin{align}
\label{eq1:lindeberg2}
\sum_{i=1}^n E(Y_{ni}^4) & = n^{-2} \sum_{i=1}^n E\left[\left\{\mathbf{d}_n^T \Sigma^{-1/2} \mathbf{U}
_{i}\right\}^4\right] \notag \\
& \leq  n^{-2} \sum_{i=1}^nE\left[ || \mathbf{d}_n||_2^4 \cdot ||\Sigma(\bbeta_0)^{-1/2}||_2^4 \cdot ||\mathbf{U}
_{i}||_2^4\right] \notag  \\
& = n^{-2} \eigen_{\max}^2\{\Sigma(\bbeta_0)^{-1} \}\sum_{i=1}^nE(||\mathbf{U}_{i}||_2^4) \notag \\
& = n^{-2} \eigen_{\max}^2\{\Sigma(\bbeta_0)^{-1} \} \sum_{i=1}^n \sum_{j=1}^{p_n} \sum_{k = 1}^{p_n} E(U_{ij}^2U_{ik}^2)  \notag \\
& = O(p_n^2/n),
\end{align}
where the first inequality is due to Cauchy-Schwarz, the second equality is due to $||\mathbf{d}_n||_2 = 1$, Condition (C4) and the definition of the spectral norm, and the last step follows from Condition  (C5). 
Therefore for any $\epsilon > 0$, 
\begin{align*}
\frac{1}{s_n^2} \sum_{i=1}^n E\left\{Y_{ni}^2 I(|Y_{ni}| > \epsilon s_n) \right\} & \leq \frac{1}{s_n^2} \sum_{i=1}^n \left\{ E(Y_{ni}^4) \right\}^{1/2} \left[ E\left\{I(|Y_{ni}| > \epsilon s_n)\right\}^2 \right]^{1/2} \\
& \leq \frac{1}{s_n^2} \left\{\sum_{i=1}^n E(Y_{ni}^4)\right\}^{1/2} \cdot \left\{\sum_{i=1}^n \Pr(|Y_{ni}|>\epsilon s_n)\right\}^{1/2} \\
& \le \frac{1}{s_n^2} O(p_n/\sqrt{n}) \cdot \left\{\sum_{i=1}^n \frac{Var(Y_{ni})}{\epsilon^2 s_n^2}\right\}^{1/2} \\
& = \frac{1}{s_n^2 \epsilon} O(p_n/\sqrt{n}) \to 0,
\end{align*}
where the third inequality follows  (\ref{eq1:lindeberg2}) and Chebyshev inequality, and last step is a consequence of $s_n^2\to 1$ and the assumption $p_n^4/n \to 0$. Thus, (\ref{eq1:lindeberg_cond}) is satisfied and consequently
\begin{align*}
n^{-1/2} \mathbf{d}_n^T \Sigma(\bbeta_0)^{-1/2} \dot{l}_{n}(\bbeta_0) & = s_n \frac{1}{s_n} \sum_{i=1}^n Y_{ni} + o_p(1) 
 \stackrel{D}{\to} N(0, 1),
\end{align*}
by the Lindeberg-Feller central limit theorem and Slutsky's theorem. This completes  the proof.  $\qquad \Box$
 
 \begin{lemma}[Consistency of Ridge Estimator]
\label{lem1:ridge}
Let 
\begin{align*}
\hat{\bbeta}_{ridge} & = \arg \min_{\bbeta} \left\{-2l_n(\bbeta) + \sum_{j=1}^{p_n} \xi_n \beta_j^2\right\}, \\
\end{align*}
be the Cox ridge estimator defined in Equation (\ref{eq1:init_ridge}). Assume that Conditions (C1) - (C5), and (C6)(i) and (C6)(iii) hold. Then 
\begin{equation}
\label{eq1:ridge_result}
||\hat{\bbeta}_{ridge} - \bbeta_0 ||_2 = O_p\left[ \sqrt{p_n}\{n^{-1/2}(1 + \xi_n b_n / \sqrt{n})\}\right]=O_p(\sqrt{p_n/n}),
\end{equation}
where $b_n$ is an upper bound of the true nonzero  $|\beta_{0j}|$'s defined in Condition (C6).
\end{lemma}

{\bf{Proof.}} Let $\alpha_n = \sqrt{p_n}(n^{-1/2} + \xi_nb_n/n)$ and $L_n (\bbeta)= -2l_n(\bbeta) + \xi_n \sum_{j=1}^{p_n}\beta_j^2$.  To prove Lemma \ref{lem1:ridge}, it is sufficient to show that for any  $\epsilon > 0$, there exists a large enough constant $K_0$ such that
\begin{equation}
\label{eq1:ridge_cond}
\Pr \left\{ \inf_{||\mathbf{u}||_2 = K_0} L_n(\bbeta_0 + \alpha_n\mathbf{u}) > L_n(\bbeta_0) \right\} \geq 1 - \epsilon,
\end{equation}
since (\ref{eq1:ridge_cond}) implies that there exists a local minimum, $\hat{\bbeta}_{ridge}$, inside the ball $\{\bbeta_0 + \alpha_n \mathbf{u}: ||\mathbf{u}||_2 \leq K_0\}$ such that $||\hat{\bbeta}_{ridge} - \bbeta_0||_2 = O_p(\alpha_n)$, with probability tending to one.  To prove (\ref{eq1:ridge_cond}), we first note 
\begin{align*}
\frac{1}{n} L_n(\bbeta_0 + \alpha_n \mathbf{u}) - \frac{1}{n}  L_n(\bbeta_0) 
& = - \frac{1}{n}  \{2l_n(\bbeta_0 + \alpha_n \mathbf{u}) -  \frac{1}{n}  2l_n(\bbeta_{0})\} +  \frac{\xi_n}{n}  \sum_{j = 1}^{p_n} \left\{  (\beta_{0j} + \alpha_n u_j)^2 -   \beta_{0j}^2\right\} \\
& = - \frac{1}{n}  \{2l_n(\bbeta_0 + \alpha_n \mathbf{u}) -  2l_n(\bbeta_{0})\} + \frac{\xi_n}{n} \sum_{j = 1}^{p_n} \left(2 \beta_{0j}\alpha_nu_j +  \alpha_n^2u_j^2 \right) \\
& \geq -\frac{1}{n}  \{2l_n(\bbeta_0 + \alpha_n \mathbf{u}) -  2l_n(\bbeta_{0})\} + \frac{2\xi_n\alpha_n}{n}  \sum_{j = 1}^{p_n}  \beta_{0j}u_j \\
& = - \frac{1}{n}  \{2l_n(\bbeta_0 + \alpha_n \mathbf{u}) -  2l_n(\bbeta_{0})\} + \frac{2\xi_n\alpha_n}{n} \sum_{j = 1}^{q_n}\beta_{0j}u_j \\
& \equiv W_1 + W_2.
\end{align*}
By Taylor expansion, we have
\begin{align*}
W_1 & = - \frac{2}{n} \alpha_n \mathbf{u}^T\dot{l}_n(\bbeta_0) - \frac{1}{n} \alpha_n^2 \mathbf{u}^T \ddot{l}_n(\bbeta^*)\mathbf{u} \\
& = W_{11} + W_{12},
\end{align*}
where $\bbeta^*$ lies between $\bbeta_0$ and $\bbeta_0 + \alpha_n \mathbf{u}$, and $\dot{l}_n(\bbeta)$ and $\ddot{l}_n(\bbeta)$ denote the first and second derivatives of $l_n(\bbeta)$, respectively. By the Cauchy-Schwartz inequality, 
\begin{align*}
W_{11} =  -\frac{2}{n} \alpha_n \mathbf{u}^T\dot{l}_n(\bbeta_0) \leq \frac{2}{n} \alpha_n || \dot{l}_n(\bbeta_0)||_2 \cdot ||\mathbf{u}||_2 =
 \frac{2}{n} \alpha_n O_p(\sqrt{np_n}) ||\mathbf{u}||_2  \leq O_p(\alpha_n^2)||\mathbf{u}||_2,
\end{align*}
where the second equality holds because $||\dot{l}_n(\bbeta_0)||_2 = O_p(\sqrt{np_n})$ from Lemma \ref{lem1:consistency} under Conditions (C1) - (C5), and the last inequality is due to $\sqrt{p_n/n} \leq \alpha_n$. By equation (A.4) of \cite{cai2005variable}, under conditions (C1)-(C5) and 
$p_n^4/n \to 0$, we have
\begin{align}
\label{eq1:uniform_hessian}
\left\|n^{-1} \ddot{l}_n(\bbeta) + \Sigma(\bbeta) \right\|_2 = o_p(p_n^{-1}),
\end{align}
in probability, uniformly in $\bbeta \in \mathcal{B}_0$. Hence
\begin{align*}
W_{12} & = -\frac{1}{n} \alpha_n^2\mathbf{u}^T\ddot{l}_n(\bbeta^*)\mathbf{u} =  \alpha_n^2\mathbf{u}^T\Sigma(\bbeta_0)\mathbf{u}\{1 + o_p(1)\}.
\end{align*}
Since $\eigen_{min}\{\Sigma(\bbeta_0)\} \geq C_1^{-1} > 0$ by Condition (C4), $W_{12}$ dominates $W_{11}$ uniformly in $||\mathbf{u}||_2 = K_0$ for a sufficiently large $K_0$. Furthermore
\begin{align*}
W_2 & \leq \frac{2\xi_n\alpha_n}{n} | \bbeta_{01}^T \mathbf{u} |   \leq \frac{2\sqrt{q_n}\xi_n\alpha_nb_n}{n} ||\mathbf{u} ||_2 = O_p(\alpha_n^2) ||\mathbf{u}||_2,
\end{align*}
where the last step follows from the fact that $\sqrt{q_n}\xi_nb_n/n < \sqrt{p_n}(n^{-1/2} + \xi_nb_n/n) = \alpha_n$. 
Therefore for a sufficiently large $K_0$, we have that $W_{12}$ dominates $W_{11}$  and $W_2$ uniformly in $||\mathbf{u}||_2 =K_0$. Since $W_{12}$ is positive, (\ref{eq1:ridge_cond}) holds and therefore $||\hat{\bbeta}_{ridge} - \bbeta_0||_2 = O_p(\alpha_n) = O_p \left[\sqrt{p_n}\{n^{-1/2}(1 + \xi_n b_n / \sqrt{n})\}\right] =O_p(\sqrt{p_n/n})$, where the last step follows from condition (C6)(iii).  
$\Box$

\begin{remark}
Let $\hat{\bbeta}_{ridge, 1}$ and $\hat{\bbeta}_{ridge, 2}$ denote the first $q_n$ and the remaining $p_n - q_n$ components of $\hat{\bbeta}_{ridge}$, respectively. 
Then,  Lemma \ref{lem1:ridge} and condition (C6) imply that for $j = 1, \ldots, q_n$ and sufficiently large $n$, $a_n/2 \leq |\hat{\beta}_{ridge, 1j} | \leq 2b_n$, where $\hat{\beta}_{ridge, 1j}$ is the $j^{th}$ component of $\hat{\bbeta}_{ridge, 1}$ and $|| \hat{\bbeta}_{ridge, 2} ||_2 = O(\sqrt{p_n/n})$.
\end{remark}

\begin{lemma}
\label{lem1:map}
Let $M_n = \max\{2/a_n, 2b_n\}$. Define $\mathcal{H}_n \equiv\{\bbeta={ \left(\bbeta_1^T, \bbeta_2^T\right)^T}: |\bbeta_1| = (|\beta_1|, \ldots, |\beta_{q_n}|)^T  \in [1/M_n, M_n]^{q_n}, 0<\| \bbeta_2 \|_2 \leq \delta_n\sqrt{p_n/n}, \}$, where $\delta_n$ is a sequence of positive real numbers satisfying $\delta_n\rightarrow \infty$ and $p_n\delta_n^2/\lambda_n\rightarrow 0$.
For any given $\bbeta \in \mathcal{H}_n$, define
\begin{equation}
\label{eq1:bar_objective}
Q_n(\btheta| \bbeta) = -2 l_n(\btheta) + \lambda_n \btheta^T D(\bbeta)\btheta,
\end{equation}
where $l_n(\btheta)$  is the $p_n$-dimensional log-partial likelihood and $D(\bbeta) = diag(\beta_1^{-2}, \ldots, \beta_{p_n}^{-2})$. 
Let  $g({\bbeta}) = \left(g_1(\bbeta)^T, g_2(\bbeta)^T \right)^T$ be a solution to $\dot{Q}_n(\btheta|\bbeta) = \mathbf{0}$, where 
\begin{equation}
\label{eq1:deriv1}
\dot{Q}_n(\btheta|\bbeta) = -2 \dot{l}_n(\btheta) + 2\lambda_n D(\bbeta)\btheta,
\end{equation}

is the derivative of $Q(\btheta|\bbeta)$ with respective to $\btheta$.
 Assume that conditions (C1) - (C6) hold. 
 Then, as $n\to \infty$, with probability tending to 1, 
\begin{itemize}
\item[(a)] 
$\sup_{\bbeta \in \mathcal{H}_n}\frac{\| g_2(\bbeta)\|_2}{\|\bbeta_2\|_2} \le  \frac{1}{K_1},
\quad \mbox{for  some constant $K_1 > 1$}$;
\item[(b)] $\left| g_1(\bbeta) \right| \in [1/M_n, M_n]^{q_n}$. 
\end{itemize}
\end{lemma}

{\bf{Proof.}}  By the first-order Taylor expansion and the definition of $g(\bbeta)$, we have 
\begin{equation}
\label{eq1:taylor_q}
\dot{Q}_n(\bbeta_0|\bbeta) = \dot{Q}_n\{g(\bbeta)|\bbeta\} + \ddot{Q}_n(\bbeta^{*}|\bbeta)\{\bbeta_0 -  g(\bbeta)\} = \ddot{Q}_n(\bbeta^{*}|\bbeta)\{\bbeta_0 -  g(\bbeta)\} ,
\end{equation}
where $\bbeta_0$ is the true parameter vector, and $\bbeta^{*}$ lies between $\bbeta_0$ and $g(\bbeta)$. Rearranging terms, we have 
\begin{equation}
\label{eq1:taylor_q2}
\ddot{Q}_n(\bbeta^{*}|\bbeta)g(\bbeta) = -\dot{Q}_n(\bbeta_0|\bbeta) + \ddot{Q}_n(\bbeta^{*}|\bbeta)\bbeta_0,
\end{equation}
which can be rewritten as
\begin{align*}
\left\{ -2\ddot{l}_n(\bbeta^{*}) + 2\lambda_n D(\bbeta) \right\} g(\bbeta) & = -\left\{ -2\dot{l}_n({\bbeta_0}) + 2 \lambda_n D(\bbeta){\bbeta_0} \right\} + \left\{-2\ddot{l}_n(\bbeta^{*}) + 2 \lambda_n D(\bbeta) \right\} \bbeta_0 \\
& = 2\dot{l}_n(\bbeta _0) - 2\ddot{l}_n(\bbeta^{*}){\bbeta_0}.
\end{align*}
Write $ H_n(\bbeta) \equiv -n^{-1} \ddot{l}_n(\bbeta)$, we have
\begin{equation}
\label{eq1:obj0a}
\left\{ H_n(\bbeta^{*}) + \frac{\lambda_n}{n}D(\bbeta)\right\} g(\bbeta)  = H_n(\bbeta^{*})\bbeta_0 +  \frac{1}{n}\dot{l}_n(\bbeta_0),
\end{equation}

which can be further written as 
\begin{equation}
\label{eq1:obj1}
\{g(\bbeta)-\bbeta_0\} + \frac{\lambda_n}{n} H_n(\bbeta^{*})^{-1} D(\bbeta)g(\bbeta) = \frac{1}{n} H_n(\bbeta^{*})^{-1} \dot{l}_n(\bbeta_0).
\end{equation}
Now we partition $H_n(\bbeta^{*})^{-1}$  into
\[ H_n(\bbeta^{*})^{-1} = \left[ \begin{array}{ll}
A & B \\
B^T & G
\end{array} 
\right] \] 
and  partition $D(\bbeta)$ into 
\[ D(\bbeta) = \left[ \begin{array}{ll}
D_1(\bbeta_1) & \mathbf{0} \\
\mathbf{0}^T & D_2(\bbeta_2)
\end{array} 
\right] \] 
where $D_1(\bbeta_1)=\mbox{diag}(\beta_1^{-2},...,\beta_{q_n}^{-2})$ and $D_2(\bbeta_2)=\mbox{diag}(\beta_{q_n+1}^{-2},...,\beta_{p_n}^{-2})$. Then (\ref{eq1:obj1}) can be re-written as
\begin{align}\label{eqnA.15}
\left( \begin{array}{c}
   g_1(\bbeta) - \bbeta_{01} \\
   g_2(\bbeta)
\end{array} \right)  + 
\frac{\lambda_n}{n}
\left( \begin{array}{l}
A D_1(\bbeta_1) g_1(\bbeta) + BD_2(\bbeta_2)g_2(\bbeta)\\
B^T D_1(\bbeta_1) g_1(\bbeta) + GD_2(\bbeta_2)g_2(\bbeta) 
\end{array} \right)
= \frac{1}{n} H_n(\bbeta^{*})^{-1} \dot{l}_n(\bbeta_0).
\end{align}
Moreover, it follows from (\ref{eq1:uniform_hessian}), condition (C4) and Lemma \ref{lem1:consistency} that
 \begin{equation}\label{eqnA.16}
 \left\| n^{-1}H_n(\bbeta^{*})^{-1} \dot{l}_n(\bbeta_0) \right\|_2 = O_p(\sqrt{p_n/n}).
 \end{equation}
  Therefore,
\begin{equation}
\label{eq1:g2eq1}
\sup_{\bbeta \in \mathcal{H}_n}\left\| g_2(\bbeta) + \frac{\lambda_n}{n}B^T D_1(\bbeta_1) g_1(\bbeta) + \frac{\lambda_n}{n}GD_2(\bbeta_2)g_2(\bbeta) 
\right\|_2 = O_p(\sqrt{p_n/n}).
\end{equation}
Furthermore, 
\begin{align*}
\left\| g(\bbeta) - \bbeta_0 \right\|_2 &= \left\| -\left\{H_n(\bbeta^{*}) +  \frac{\lambda_n}{n} D(\bbeta)\right\}^{-1}
\left\{ \frac{\lambda_n}{n} D(\bbeta) \bbeta_0 - \frac{1}{n}\dot{l}_n(\bbeta_0) \right\}\right\|_2\\
 & \leq  \left\| \left\{H_n(\bbeta^{*}) \right\}^{-1} 
\left\{ \frac{\lambda_n}{n} D(\bbeta) \bbeta_0 - \frac{1}{n}\dot{l}_n(\bbeta_0) \right\} \right\|_2 \\
& \leq \left\| \left\{H_n(\bbeta^{*}) \right\}^{-1} \right\|_2 \cdot \left\{
 \left\|\frac{\lambda_n}{n} D_1(\bbeta_1) \bbeta_{01}\right\|_2 + 
 \left\|\frac{1}{n}\dot{l}_n(\bbeta_0) \right\|_2  \right\}\\
& = O_p(1) \left\{ O(n^{-1}\lambda_n M_n^3\sqrt{q_n} ) + O_p(\sqrt{p_n/n})  \right\} \\
& = O_p(\sqrt{p_n/n}),
\end{align*}
where the first equality follows from (\ref{eq1:obj0a}) and  the fourth step follows from (\ref{eq1:uniform_hessian}), condition (C4), $\left\| n^{-1}\lambda_nD_1(\bbeta_1) \beta_{01} \right\|_2 = 
O(n^{-1}\lambda_n M_n^3 \sqrt{q_n} ) $, and $\left\| n^{-1} \dot{l}_n(\bbeta_0) \right\|_2 = O_p(\sqrt{p_n/n})$, and the last step holds  since $n^{-1}\lambda_n M_n^3\sqrt{q_n} = o(1/\sqrt{n})$ under condition (C6).
Hence,
\begin{align} \label{A.15}
\left\| g(\bbeta) \right\|_2  \leq \left\| \bbeta_0 \right\|_2 +\left\| g(\bbeta) - \bbeta_0 \right\|_2 =   O_p(M_n\sqrt{q_n}).
\end{align}
Also note that $\left\| B  \right\|_2 =O_p(1)$ since
$\left\| B B^T \right\|_2    \le \left\|A^2 + BB^T \right\|_2 + \left\| A ^2\right\|_2 \le 2\left\|A^2 + BB^T \right\|_2
 \leq 2\left\| H_n(\bbeta^{\ast})^{-2} \right\|_2 =O_p(1)$.
This, combined with (\ref{A.15}), implies that
\begin{equation}
\label{eq1:g2eq2}
\sup_{\bbeta \in \mathcal{H}_n}\left\| \frac{\lambda_n}{n}B^T  D_1(\bbeta_1) g_1(\bbeta) \right\|_2   \leq \frac{\lambda_n}{n} \sup_{\bbeta \in \mathcal{H}_n} \left\| B \right\|_2 \left\| D_1({\bbeta_1})\right\|_2  \left\| g_1(\bbeta) \right\|_2   
= O_p\left( \frac{\lambda_nM_n^{3} \sqrt{q_n}}{n}\right)=o(1/\sqrt{n}).
\end{equation}
It then follows from (\ref{eq1:g2eq1}) and (\ref{eq1:g2eq2}) that 
\begin{align*}
\sup_{\bbeta \in \mathcal{H}_n}\left\| g_2(\bbeta) + \frac{\lambda_n}{n} GD_2(\bbeta_2)g_2(\bbeta) \right\|_2 & \le O_p(\sqrt{p_n/n}) + o(1/\sqrt{n})
 = O_p(\sqrt{p_n/n}).
\end{align*}
Since $G$ is positive definite and symmetric with probability tending to one, by the spectral decomposition theorem, $G = \sum_{i=1}^{p_n-q_n} r_{2i}\mathbf{u}_{2i}\mathbf{u}_{2i}^T$, where $r_{2i}$ and $\mathbf{u}_{2i}$ are the eigenvalues and eigenvectors of $G$, respectively. Now with probability tending to one,
\begin{align}
\label{eq1:m2bound1}
\frac{\lambda_n}{n} \left\| G D_2(\bbeta_2)g_2(\bbeta) \right\|_2 & = \frac{\lambda_n}{n} \left\|  \left( \sum_{i=1}^{p_n-q_n} r_{2i}\mathbf{u}_{2i}\mathbf{u}_{2i}^T \right) D_2(\bbeta_2)g_2(\bbeta) \right\|_2 \notag \\
& \geq  \frac{\lambda_n}{n} \left\| \frac{1}{C_1}  \left( \sum_{i=1}^{p_n-q_n} \mathbf{u}_{2i} \mathbf{u}_{2i}^T \right) D_2(\bbeta_2)g_2(\bbeta)  \right\|_2 \notag \\
& \geq \frac{1}{C_1} \left\| \frac{\lambda_n}{n} D_2(\bbeta_2) g_2(\bbeta) \right\|_2,
\end{align}
where the first inequality is due to (\ref{eq1:uniform_hessian}) and condition (C4) since we can assume that for all $i = 1, \ldots, p_n - q_n$,   $r_{2i} \in (1/C_1, C_1)$ for some $C_1 > 1$ with probability tending to one.
Therefore with probability tending to one,
\begin{align}
\label{eq1:m2bound4}
\frac{1}{C_1} \left\| \frac{\lambda_n}{n} D_2(\bbeta_2)g_2(\bbeta) \right\|_2 - \left\| g_2(\bbeta) \right\|_2 \leq \left\| g_2(\bbeta) + \frac{\lambda_n}{n} GD_2(\bbeta_2)g_2(\bbeta) \right\|_2  \leq \delta_n \sqrt{p_n/n},
\end{align}
where $\delta_n$ diverges to $\infty$.
Let $\mathbf{m}_{g_2(\bbeta)/\bbeta_2} = (g_2(\beta_{q_n+1})/\beta_{q_n+1}, \ldots, g_2(\beta_{p_n})/\beta_{p_n})^T$. Because
$||\bbeta_2||_2 \leq \delta_n \sqrt{p_n/n}$, we have
\begin{align}
\label{eq1:m2bound2}
\frac{1}{C_1} \left\|\frac{\lambda_n}{n} D_2(\bbeta_2) g_2(\bbeta) \right\|_2  = \frac{1}{C_1} \frac{\lambda_n}{n} \left\| D_2(\bbeta_2)^{1/2}\mathbf{m}_{g_2(\bbeta)/\bbeta_2} \right\|_2 
\geq  \frac{1}{C_1} \frac{\lambda_n}{n} \frac{\sqrt{n}}{\delta_n \sqrt{p_n}} \left\| \mathbf{m}_{g_2(\bbeta)/\bbeta_2} \right\|_2,
\end{align}
and
\begin{equation}
\label{eq1:m2bound3}
\left\| g_2(\bbeta) \right\|_2 = \left\|  D_2(\bbeta_2)^{-1/2}\mathbf{m}_{g_2(\bbeta)/\bbeta_2} \right\|_2 \leq  \left\|  D_2(\bbeta_2)^{-1/2} \right\|_2 \cdot \left\| \mathbf{m}_{g_2(\bbeta)/\bbeta_2} \right\|_2  \leq \frac{\delta_n \sqrt{p_n}}{\sqrt{n}} \left\| \mathbf{m}_{g_2(\bbeta)/\bbeta_2} \right\|_2.
\end{equation}
 Hence it follows from (\ref{eq1:m2bound4}), (\ref{eq1:m2bound2}), and (\ref{eq1:m2bound3}) that with probability tending to one,
\begin{align*}
\frac{1}{C_1} \frac{\lambda_n}{n} \frac{\sqrt{n}}{\delta_n \sqrt{p_n}} \left\| \mathbf{m}_{g_2(\bbeta)/\bbeta_2} \right\|_2 - \frac{\delta_n \sqrt{p_n}}{\sqrt{n}} \left\| \mathbf{m}_{g_2(\bbeta)/\bbeta_2} \right\|_2 \leq \delta_n \sqrt{p_n/n}.
\end{align*}
This implies that with probability tending to one,
\begin{equation}
\label{m2bound}
\left\| \mathbf{m}_{g_2(\bbeta)/\bbeta_2} \right\|_2 \leq \frac{1}{\lambda_n/(C_1p_n\delta_n^2) - 1 }< \frac{1}{K_1},
\end{equation}
for some constant $K_1 > 1$ provided that $\lambda_n/(p_n\delta_n^2) \to \infty$ as $n \to \infty$. Now from (\ref{m2bound}), we have 
\begin{align}\label{m2bound1}
\left\| g_2(\bbeta) \right\|_2 \leq \left\| \mathbf{m}_{g_2(\bbeta)/\bbeta_2} \right\|_2  \max_{q_n+1 \leq j \leq  p_n} |\beta_j| \leq \left\|  \mathbf{m}_{g_2(\bbeta)/\bbeta_2} \right\|_2  \left\| \bbeta_2\right\|_2 \leq \frac{1}{K_1} \left\|\bbeta_2 \right\|_2,
\end{align}
with probability tending to one. 
Thus
\begin{align*}
\Pr \left( \sup_{\bbeta \in \mathcal{H}_n} \frac{ \left\| g_2(\bbeta) \right\|_2  }{  \left\| \bbeta_2 \right\|_2 } < \frac{1}{K_1} \right) \to 1 \hspace{.2in} \mbox{as $n \to \infty$}
\end{align*}
and (a) is proved.

To prove part (b),  we first note from (\ref{m2bound1}) that as $n\to\infty$,
$
\Pr(\left\| \mathbf{m}_{g_2(\bbeta)/\bbeta_2} \right\|_2 \le  \delta_n\sqrt{p_n/n}) \to 1.
$
Therefore it is sufficient to show that for any $\bbeta \in \mathcal{H}_n$, 
$\left| g_1(\bbeta) \right| \in [1/M_n, M_n]^{q_n}$
 with probability tending to 1.
 By  (\ref{eqnA.15}) and (\ref{eqnA.16}), we have 
\begin{align}
\label{g1eq1}
\sup_{\bbeta \in \mathcal{H}_n}\left\| (g_1(\bbeta) - \bbeta_{01} ) + \frac{\lambda_n}{n}A D_1(\bbeta_1) g_1(\bbeta) + \frac{\lambda_n}{n}BD_2(\bbeta_2)g_2(\bbeta) 
\right\|_2=O_p(\sqrt{p_n/n}).
\end{align}
Similar to (\ref{eq1:g2eq2}), it can be shown that 
\begin{align}\label{g1eq1a}
\sup_{\bbeta \in \mathcal{H}_n}\left\| \frac{\lambda_n}{n}A D_1(\bbeta_1) g_1(\bbeta) \right\|_2 = O_p\left( \frac{\lambda_nM_n^{3}\sqrt{q_n}}{n}\right) = o_p(1/\sqrt{n}),
\end{align}
where the last equality holds trivially under condition (C6).
Furthermore,  with probability tending to one,
\begin{align}\label{g1eq1b}
\sup_{\bbeta \in \mathcal{H}_n}\left\| \frac{\lambda_n}{n}BD_2(\bbeta_2)g_2(\bbeta) 
\right\|_2 \leq \frac{\lambda_n}{n} \sup_{\bbeta \in \mathcal{H}_n} \left\| B \right\|_2 \cdot \left\| D_2(\bbeta_2)g_2(\bbeta) \right\|_2   \leq  \frac{\lambda_n}{n} \sqrt{2K_3} \left( \delta_n \sqrt{\frac{p_n}{n}} \right)^2,
\end{align}
for some $K_3>0$, since $||g_2(\bbeta)|| \le \delta_n \sqrt{p_n/n}$, $||B||_2 =O_p(1)$
and $\left\| D_2(\bbeta_2) \right\|_2 \le \delta_n \sqrt{p_n/n}$. Therefore,  combing (\ref{g1eq1}),  (\ref{g1eq1a}) and (\ref{g1eq1b}) gives
\begin{align*}
\sup_{\bbeta \in \mathcal{H}_n} \left\| g_1(\bbeta) - \bbeta_{01} \right\|_2 \leq \frac{\lambda_n}{n} \sqrt{2K_3} \left( \delta_n \sqrt{\frac{p_n}{n}} \right)^2 + \frac{\delta_n\sqrt{p_n}}{\sqrt{n}},
\end{align*}
with probability tending to one. Because  $\lambda_n/n \to 0$ and $\delta_n \sqrt{p_n/n} =\sqrt{p_n\delta_n^2/\lambda_n}  \sqrt{{\lambda_n}/{n}} \to 0$  as $n \to \infty$, we have   
$\Pr(\left| g_1(\bbeta) \right| \in [1/M_n, M_n]^{q_n}) \to 1$. This completes the proof of part (b). $\Box$

\begin{lemma}
\label{lem1:asymptotic}
Let $\bbeta_1$ be the first $q_n$ components of $\bbeta$. Define  $f(\bbeta_1) = \arg \min_{\btheta_1} \{ Q_{n1}(\btheta_1|\bbeta_1) \}$,
where
$
Q_{n1}(\btheta_1|\bbeta_1) = -2 l_{n1}(\btheta_1) + \lambda_n \btheta_1^T D_1(\bbeta_1)\btheta_1,
$
is a weighted $L_2$-penalized -2log-partial likelihood for the oracle model of model size $q_n$, and
$D_1(\bbeta_1) = diag(\beta_1^{-2}, \beta_2^{-2}, \ldots, \beta_{q_n}^{-2})$. Assume that conditions (C1) - (C6) hold. Then with probability tending to one,
\begin{itemize}
\item[(a)] $f(\bbeta_1)$ is a contraction mapping from $[1/M_n, M_n]^{q_n}$ to itself;
\item[(b)] $\sqrt{n}\mathbf{b}_n^T  \Sigma(\bbeta_0)_{11}^{1/2}(\hat{\bbeta}_1^{\circ} - \bbeta_{01}) \stackrel{D}{\to} N(0, 1)$, for any $q_n$-dimensional vector $\mathbf{b}_n$ such that $\mathbf{b}_n^T \mathbf{b}_n = 1$ and  where $\hat{\bbeta}_1^{\circ}$ is the unique fixed point of $f(\bbeta_1)$ and $\Sigma(\bbeta_0)_{11}$ is the first $q_n \times q_n$ submatrix of $\Sigma(\bbeta_0)$.  
\end{itemize}
\end{lemma}

{\bf{Proof:}} (a) First we show that $f(\cdot)$ is a mapping from $[1/M_n, M_n]^{q_n}$ to itself with probability tending to one. Again through a first order Taylor expansion, we have 
 \begin{equation}
 \label{eq1:scoref1}
 \{f(\bbeta_1) - \bbeta_{01}\} + \frac{\lambda_n}{n} H_{n1}(\bbeta_1^{*})^{-1}D_1(\bbeta_1)f(\bbeta_1)
 = \frac{1}{n} H_{n1}(\bbeta_1^{*})^{-1} \dot{l}_{n1}(\bbeta_{01}),
 \end{equation}
 where $H_{n1}(\bbeta_1^{*}) = -n^{-1}\ddot{l}_{n1}(\bbeta_1^{*})$ exists and is invertible for $\bbeta_1^{*}$ between $\bbeta_{01}$ and $f({\bbeta_1})$. We have 
\begin{align*}
\label{eq1:f1eq1}
\sup_{|\bbeta_1| \in [1/M_n, M_n]^{q_n}} \left\| f(\bbeta_1) - \bbeta_{01} + \frac{\lambda_n}{n}H_{n1}(\bbeta_1^{*})^{-1}D_1(\bbeta_1)f(\bbeta_1) \right\|_2 = O_p(\sqrt{q_n/n}),
 \end{align*}
where the right-hand side follows in the same fashion as (\ref{eq1:g2eq1}). Similar to (\ref{eq1:g2eq2}) we have
\begin{align*}
\sup_{|\bbeta_1| \in [1/M_0, M_0]^{q_n}} \left\| \frac{\lambda_n}{n}H_{n1}(\bbeta_1^{*})^{-1}D_1(\bbeta_1)f(\bbeta_1) \right\|_2 & = O_p\left(\frac{\lambda_n M_n^3}{\sqrt{n}} \sqrt{\frac{q_n}{n}} \right) = o_p\left( 1/\sqrt{n} \right).  
\end{align*}

Therefore, with probability tending to one
\begin{equation}
\label{eq1:fbound}
\sup_{|\bbeta_1| \in [1/M_n, M_n]^{q_n}} \left\| f(\bbeta_1) - \bbeta_{01} \right\|_2 \leq \delta_n \sqrt{q_n/n},
\end{equation}
where $\delta_n$ is a sequence of real numbers diverging to $\infty$ and satisfies $\delta_n \sqrt{p_n/n} \to 0$.
As a result, we have
\begin{align*}
\Pr( f(\bbeta_1) \in [1/M_n, M_n]^{q_n}) \to 1
\end{align*}
as $n \to \infty$. Hence $f(\cdot)$ is a mapping from the region $[1/M_n, M_n]^{q_n}$ to itself. To prove that $f(\cdot)$ is a contraction mapping,  we  need to further show  that
\begin{equation}
\label{eq1:contractresult}
\sup_{|\bbeta_1| \in [1/M_n, M_n]^{q_n}} \left\| \dot{f}(\bbeta_1) \right\|_2 = o_p(1).
\end{equation}
Since $f(\bbeta_1)$ is a solution to $\dot{Q}_{n1}(\btheta_1|\bbeta_1)= 0$, we have
\begin{equation}
\label{score2}
-\frac{1}{n}\dot{l}_{n1}(f(\bbeta_1)) = -\frac{\lambda_n}{n} D_1(\bbeta_1) f(\bbeta_1).
\end{equation}
Taking the derivative of (\ref{score2}) with respect to $\bbeta_1^T$ and rearranging terms, we obtain
\begin{align}\label{A.32a}
\left\{ H_{n1}(f(\bbeta_1))  + \frac{\lambda_n}{n} D_1(\bbeta_1) \right\} \dot{f}(\bbeta_1) =  \frac{2\lambda_n}{n} diag\{f_1(\bbeta_1)/\beta_1^3, \ldots, f_{q_n}(\bbeta_1)/\beta_{q_n}^3 \}.
\end{align}
With probability tending to one, we have
\begin{align*}
\sup_{|\bbeta_1| \in [1/M_n, M_n]^{q_n}}  \frac{2 \lambda_n}{n} \left\| diag\{f_1(\bbeta_1)/\beta_1^3, \ldots, f_{q_n}(\bbeta_1)/\beta_{q_n}^3 \} \right\|_2 = O_p \left( \frac{\lambda_nM_n^4}{n} \right) = o_p(1),
\end{align*}
where the last step follows from condition (C6). This, combined with (\ref{A.32a}) implies that
\begin{align}
\label{gradfpt1}
\sup_{|\bbeta_1| \in [1/M_n, M_n]^{q_n}} \left\|  \left\{ H_{n1}(f(\bbeta_1)) + \frac{\lambda_n}{n} D_1(\bbeta_1) \right\} \dot{f}(\bbeta_1)\right\|_2 = o_p(1).
\end{align}
Now, it can be shown that probability tending to one, 
\begin{align*}
\left\| H_{n1}(f(\bbeta_1))\dot{f}(\bbeta_1) \right\|_2 & \geq \left\|\dot{f}(\bbeta_1) \right\|_2  \cdot \left\|H_{n1}(f(\bbeta_1))^{-1} \right\|_2^{-1} \geq \frac{1}{K_2} \left\| \dot{f}(\bbeta_1) \right\|_2,
\end{align*}
for some $K_2>0$, and that
\begin{align*}
\frac{\lambda_n}{n} \left\| D_1(\bbeta_1) \dot{f}(\bbeta_1) \right\|_2&  \geq \frac{\lambda_n}{n} \left\|\dot{f}(\bbeta_1) \right\|_2  \left\|D_1(\bbeta_1)^{-1}\right\|_2^{-1}
 \geq \frac{\lambda_n}{n} \frac{1}{M_n^2} \left\|\dot{f}(\bbeta_1)\right\|_2. 
\end{align*}
Therefore, combining the above two inequalities with (\ref{A.32a}) and (\ref{gradfpt1}) gives
\begin{align*}
\left( \frac{1}{K_2} - \frac{\lambda_n}{nM_n^2} \right) \sup_{|\bbeta_1| \in [1/M_n, M_n]^{q_n}} \left\|\dot{f}(\bbeta_1)\right\|_2 
= o_p(1).
\end{align*}
This, together with the fact that $\frac{\lambda_n}{n} \frac{1}{M_n^2}=o(1)$, implies that 
 (\ref{eq1:contractresult}) holds. Therefore, with probability tending to one, $f(\cdot)$ is a contraction mapping and consequently has a unique fixed point, say $\hat{\bbeta}_1^{\circ}$, such that $\hat{\bbeta}_1^{\circ} = f( \hat{\bbeta}_1^{\circ})$.
   
We next prove part (b). By (\ref{eq1:scoref1}) we have
\begin{align*}
f(\bbeta_1) = \left\{ H_{n1}(\bbeta_1^*) + \frac{\lambda_n}{n}D_1(\bbeta_1) \right\}^{-1}\left\{ H_{n1}(\bbeta_1^*)\bbeta_{01} + \frac{1}{n}\dot{l}_{n1}(\bbeta_{01}) \right\}.
\end{align*}

Now,
\begin{align}
\sqrt{n} \mathbf{b}_n^T \Sigma(\bbeta_0)_{11}^{1/2}(\hat{\bbeta}_1^{\circ} - \bbeta_{01}) & = \sqrt{n} \mathbf{b}_n^T \Sigma(\bbeta_0)_{11}^{1/2}  \left[ \left\{ H_{n1}(\bbeta_1^*) + \frac{\lambda_n}{n}D_1(\hat{\bbeta}_1^{\circ}) \right\}^{-1} H_{n1}(\bbeta_1^*) - I_{q_n} \right] \bbeta_{01} \notag \\
& + \sqrt{n} \mathbf{b}_n^T \Sigma(\bbeta_0)_{11}^{1/2} \left[ \left\{ H_{n1}(\bbeta_1^*) + \frac{\lambda_n}{n}D_1(\hat{\bbeta}_1^{\circ}) \right\}^{-1} \frac{1}{n} \dot{l}_{n1}(\bbeta_{01}) \right] \notag \\
& = I_1 + I_2. \label{ni0}
\end{align}
Note that for any two conformable invertible matrices $\Phi$ and $\Psi$, we have
\begin{align*}
(\Phi + \Psi)^{-1} = \Phi^{-1} - \Phi^{-1}\Psi(\Phi + \Psi)^{-1},
\end{align*}
Thus we can rewrite $I_1$ as
\begin{align*}
I_1 & = \sqrt{n} \mathbf{b}_n^T \Sigma(\bbeta_0)_{11}^{1/2} \left[ \left\{ H_{n1}(\bbeta_1^*) + \frac{\lambda_n}{n}D_1(\hat{\bbeta}_1^{\circ}) \right\}^{-1} H_{n1}(\bbeta_1^*) - I_{q_n} \right] \bbeta_{01}  \notag \\
& = -\frac{\lambda_n}{\sqrt{n}} \mathbf{b}_n^T \Sigma(\bbeta_0)_{11}^{1/2} H_{n1}(\bbeta_1^*)^{-1}D_1(\hat{\bbeta}_1^{\circ}) \left\{ H_{n1}(\bbeta_1^*) + \frac{\lambda_n}{n}D_1(\hat{\bbeta}_1^{\circ}) \right\}^{-1} H_{n1}(\bbeta_1^*)\bbeta_{01}.
\end{align*}
Moreoever
\begin{align} \notag
\left\| I_1 \right\|_2 & \leq \frac {\lambda_n}{\sqrt{n}} \left\|\Sigma(\bbeta_0)_{11}^{1/2} \right\|_2 \left\| H_{n1}(\bbeta_1^*)^{-1} \right\|_2 \left\|D_1(\hat{\bbeta}_1^{\circ})\right\|_2
 \left\| \left\{ H_{n1}(\bbeta_1^*) + \frac{\lambda_n}{n}D_1(\hat{\bbeta}_1^{\circ}) \right\}^{-1}\right\|_2 \left\| H_{n1}(\bbeta_1^*) \right\|_2 \left\| \bbeta_{01} \right\|_2 \\
 & = \frac{\lambda_n}{\sqrt{n}} \cdot O(1) \cdot O_p(1) \cdot M_n^2 \cdot O_p(1) \cdot O_p(1) \cdot M_n\sqrt{q_n} \notag \\
& = O_p(\lambda_n M_n^3\sqrt{q_n}/\sqrt{n}) = o_p(1), \label{ni1}
\end{align}
where the first equality follows from (\ref{eq1:uniform_hessian}) and condition (C4), and the last equality is a consequence of condition (C6).
Similarly, we can rewrite $I_2$ as
\begin{align}
I_2 & = \sqrt{n}\mathbf{b}_n^T \Sigma(\bbeta_0)_{11}^{1/2} \left[ \left\{ H_{n1}(\bbeta_1^*) + \frac{\lambda_n}{n}D_1(\hat{\bbeta}_1^{\circ}) \right\}^{-1} \frac{1}{n} \dot{l}_{n1}(\bbeta_{01}) \right]  \notag \\
& =  \mathbf{b}_n^T \Sigma(\bbeta_0)_{11}^{1/2}H_{n1}(\bbeta_1^*)^{-1}\frac{1}{\sqrt{n}}\dot{l}_{n1}(\bbeta_{01})  \notag\\
& - \frac{\lambda_n}{\sqrt{n}} \mathbf{b}_n^T \Sigma(\bbeta_0)_{11}^{1/2}H_{n1}(\bbeta_1^*)^{-1}D_1(\hat{\bbeta}_1^{\circ}) \left\{ H_{n1}(\bbeta_1^*)^{-1} + \frac{\lambda_n}{n}D_1(\hat{\bbeta}_1^{\circ}) \right\}^{-1}\frac{1}{n}\dot{l}_{n1}(\bbeta_{01}) \notag \\
& =  \mathbf{b}_n^T \Sigma(\bbeta_0)_{11}^{1/2}H_{n1}(\bbeta_1^*)^{-1}\frac{1}{\sqrt{n}}\dot{l} _{n1}(\bbeta_{01}) + o_p(1).\label{i2}
\end{align}
We now establish the asymptotic normality of $n^{-1/2} \mathbf{b}_n^T \Sigma(\bbeta_0)_{11}^{1/2}H_{n1}(\bbeta_1^*)^{-1}\dot{l}_{n1}(\bbeta_{01})$, which will be derived in a similar fashion to Lemma \ref{lem1:consistency}.
By (\ref{eq1:uniform_hessian}), (\ref{eq1:fbound}), and the continuity of $\Sigma(\bbeta_0)$, we can deduce that $H_{n1}(\bbeta^*) = \Sigma(\bbeta_0)_{11} + o_p(1)$, where $\Sigma(\bbeta_0)_{11} = \Sigma(\bbeta_0)_{11}$ is the first $q_n \times q_n$ submatrix of $\Sigma(\bbeta_0)$. This, together with  (\ref{U1}) and (\ref{i2}), implies that
\begin{align}
I_2& =
n^{-1/2} \sum_{i=1}^n  \mathbf{b}_n^T \Sigma(\bbeta_0)_{11}^{1/2} H_{n1}(\bbeta_1^*)^{-1}\mathbf{U}
_{i1} + o_p(1) \notag \\
& = n^{-1/2} \sum_{i=1}^n  \mathbf{b}_n^T \Sigma(\bbeta_0)_{11}^{-1/2} \mathbf{U}
_{i1} 
 + \left\{ n^{-1/2} \sum_{i=1}^n \mathbf{b}_n^T \Sigma(\bbeta_0)_{11}^{1/2} \mathbf{U}
_{i1}\right\} o_p(1) + o_p(1) \notag \\
& = I_{21} + I_{22} \cdot o_p(1)  + o_p(1), \label{n0}
\end{align}
where $\mathbf{U}_{i1}$ consists of the first $q_n$ components of $\mathbf{U}_i$. 
Letting $Y_{ni} =  n^{-1/2}  \mathbf{b}_n^T \Sigma(\bbeta_0)_{11}^{-1/2} \mathbf{U}
_{i1} $,  then
 \begin{align*}
s_n^2 = \sum_{i=1}^n Var(Y_{ni}) & = \frac{1}{n} \sum_{i=1}^n  \mathbf{b}_n^T \Sigma(\bbeta_0)_{11}^{-1/2} Var(\mathbf{U}
_{i1})\Sigma(\bbeta_0)_{11}^{-1/2}\mathbf{b}_n  \\
& = \mathbf{b}_n^T \Sigma(\bbeta_0)_{11}^{-1/2} \left\{ \frac{1}{n} \sum_{i=1}^n  Var(\mathbf{U}
_{i1}) \right\} \Sigma(\bbeta_0)_{11}^{-1/2}\mathbf{b}_n \to 1.
\end{align*}
To prove the asymptotic normality of $I_{21}$, we need to verify the Lindeberg condition: for all $\epsilon > 0$,
\begin{align}
\label{eq1:i21_lindeberg_cond}
\frac{1}{s_n^2} \sum_{i=1}^n E\{Y_{ni}^2 I(|Y_{ni}| \geq \epsilon s_n)\}  \to 0,
\end{align}
as $n \to \infty$. Note that
\begin{align}
\label{eq1:i21_lindeberg2}
\sum_{i=1}^n E(Y_{ni}^4) & = n^{-2} \sum_{i=1}^n E\left[\left\{\mathbf{b}_n^T \Sigma(\bbeta_0)_{11}^{-1/2} \mathbf{U}
_{i1}\right\}^4\right] \notag \\
& \leq  n^{-2} \sum_{i=1}^nE\left[ || \mathbf{b}_n||_2^4 \cdot ||\Sigma(\bbeta_0)_{11}^{-1/2}||_2^4 \cdot ||\mathbf{U}
_{i1}||_2^4\right] \notag  \\
& = n^{-2} \eigen_{\max}^2\{\Sigma(\bbeta_0)^{-1} \}\sum_{i=1}^nE(||\mathbf{U}_{i1}||_2^4) \notag \\
& = n^{-2} \eigen_{\max}^2\{\Sigma(\bbeta_0)^{-1} \} \sum_{i=1}^n \sum_{j=1}^{p_n} \sum_{k = 1}^{p_n} E(U_{ij}^2U_{ik}^2)  \notag \\
& = O(p_n^2/n),
\end{align}
where the first inequality is due to Cauchy-Schwarz, the second equality is due to $||\mathbf{b}_n||_2 = 1$ and the last step follows from conditions (C4) and (C5). 
Therefore for any $\epsilon > 0$, 
\begin{align*}
\frac{1}{s_n^2} \sum_{i=1}^n E\left\{Y_{ni}^2 I(|Y_{ni}| > \epsilon s_n) \right\} & \leq \frac{1}{s_n^2} \sum_{i=1}^n \left\{ E(Y_{ni}^4) \right\}^{1/2} \left[ E\left\{I(|Y_{ni}| > \epsilon s_n)\right\}^2 \right]^{1/2} \\
& \leq \frac{1}{s_n^2} \left\{\sum_{i=1}^n E(Y_{ni}^4)\right\}^{1/2} \cdot \left\{\sum_{i=1}^n \Pr(|Y_{ni}|>\epsilon s_n)\right\}^{1/2} \\
& \leq \frac{1}{s_n^2} \left\{\sum_{i=1}^n E(Y_{ni}^4)\right\}^{1/2} \cdot \left\{\sum_{i=1}^n \frac{Var(Y_{ni})}{\epsilon^2 s_n^2}\right\}^{1/2} \\
& =  \frac{1}{s_n^2} \left\{O(p_n^2/n)\right\}^{1/2} \frac{1}{\epsilon}\to 0.
\end{align*}
Thus, (\ref{eq1:i21_lindeberg_cond}) is satisfied and
by the Lindeberg-Feller central limit theorem and Slutsky's theorem
\begin{align}\label{n1}
I_{21} = 
s_n \left( \frac{1}{s_n} \sum_{i=1}^n Y_{ni}  \right)  \stackrel{D}{\to} N(0, 1).
\end{align}
Similarly, it can be shown that as $n\to\infty$,
\begin{align}\label{n2}
\frac{I_{22} }{\sqrt{\mathbf{b}_n^T \Sigma(\bbeta_0)_{11}^{2}\mathbf{b}_n} } \stackrel{D}{\to} N\left(0, 1\right).
\end{align}
since $\left\| \left\{\mathbf{b}_n^T \Sigma(\bbeta_0)_{11}^{2}\mathbf{b}_n + o(1)\right\}^{-1} \right\|_2 = O(1)$.
Therefore $I_{22} = O_p(1)$ and by Slutsky's theorem,
\begin{align*}
n^{-1/2}  \mathbf{b}_n^T \Sigma(\bbeta_0)_{11}^{1/2} H_{n1}(\bbeta_1^*)^{-1}\dot{l}_{n1}(\bbeta_{01}) 
& =n^{-1/2} \sum_{i=1}^n  \mathbf{b}_n^T \Sigma(\bbeta_0)_{11}^{-1/2} \mathbf{U}
_{i1} \notag \\ 
& + \left\{ n^{-1/2} \sum_{i=1}^n \mathbf{b}_n^T \Sigma(\bbeta_0)_{11}^{1/2} \mathbf{U}
_{i1}\right\} o_p(1) + o_p(1) \notag \\
& = I_{21} + I_{22} \cdot o_p(1) + o_p(1) \\
& \stackrel{D}{\to} N(0, 1).  
\end{align*}
Hence, combining (\ref{ni0}), (\ref{ni1}), (\ref{n0}),  (\ref{n1}) and  (\ref{n2}) gives
\begin{align*}
\sqrt{n}\mathbf{b}_n^T \Sigma(\bbeta_0)_{11}^{1/2}(\hat{\bbeta}_1^{\circ} - \bbeta_{01}) \stackrel{D}{\to} N(0, 1),
\end{align*}
which proves part (b). $\Box$ \\

{\bf{Proof of Theorem \ref{th1:oracle}.}} 
Part (a) of the theorem follows immediately from part (a) of Lemma \ref{lem1:map}.
Part (b) of the theorem will follow from part (b) Lemma \ref{lem1:asymptotic} and the following
\begin{equation}
\label{eq1:nzero_est}
\Pr \left(\lim_{k \to \infty} \left\| g_1(\bbeta^{(k)}) - \hat{\bbeta}_1^{\circ} \right\|_2 = 0  \right) \to 1,
\end{equation}
where $\hat{\bbeta}_1^{\circ}$ is the fixed point of $f(\bbeta_1)$ defined in Lemma \ref{lem1:asymptotic}.
Note that $g(\bbeta)$ is a solution to 
\begin{equation}
\label{eq1:scorerewrite}
-\frac{1}{n}D(\bbeta)^{-1}\dot{l}_n(\btheta) + \frac{1}{n}\lambda_n \btheta = \mathbf{0},
\end{equation}
where $D(\bbeta)^{-1} = diag\{ \beta_1^2, \ldots, \beta_{q_n}^2, \beta_{q_n+1}^2, \ldots, \beta_{p_n}^2\}$.
It is easy to see from (\ref{eq1:scorerewrite})  that
\begin{align*}
\lim_{\bbeta_2 \to 0} g_2(\bbeta) = \mathbf{0}_{p_n - q_n}.
\end{align*}
This, combined with (\ref{eq1:scorerewrite}), implies that for any $\bbeta_1$
\begin{align*}
\lim_{\bbeta_2 \to 0} g_1(\bbeta) = f(\bbeta_1).
\end{align*}
Hence, $g(\cdot)$ is continuous  and thus uniform continuous on the compact set $\bbeta \in \mathcal{H}_n$.
Hence as $k\to \infty$, 
\begin{align}\label{omega}
\omega_k \equiv \sup_{|g_1(\bbeta)| \in [1/M_n, M_n]^{q_n}} \left\|  g_1( \bbeta_1, \hat{\bbeta}_2^{(k)}) -f(\bbeta_1) \right\|_2 \to 0,
\end{align}
with probability tending to one. 
Furthermore,
\begin{align}\label{eq1:b1converge}
 \left\| \hat{\bbeta}_1^{(k+1)} - \hat{\bbeta}_1^{\circ} \right\|_2 & \leq \left\| g_1(\hat{\bbeta}^{(k)}) - f(\hat{\bbeta}_1^{(k)}) \right\|_2 + \left\|  f(\hat{\bbeta}_1^{(k)})  - \hat{\bbeta}_1^{\circ} \right\|_2
 \le \omega_k + \frac{1}{K_4} \left\| \hat{\bbeta}_1^{(k)} - \hat{\bbeta}_1^{\circ} \right\|_2,
\end{align}
for some $K_4 > 1$, where the last inequality follows from (\ref{eq1:contractresult}) and the definition of $\omega_k$.
Denote by $a_k =  \left\| \hat{\bbeta}_1^{(k)} -\hat{\bbeta}_1^{\circ} \right\|_2 $, we can rewrite (\ref{eq1:b1converge}) as
\begin{align*}
a_{k+1} \leq \frac{1}{K_4}a_{k} + \omega_k.
\end{align*}
By (\ref{omega}), for any $\epsilon > 0$, there exists an $N > 0$ such that  $\omega_k < \epsilon$ for all $k > N$. Therefore for $k > N$,  
\begin{align*}
a_{k+1} & \leq \frac{1}{K_4}a_{k} + \omega_k \\
& \leq \frac{a_{k-1}}{K_4^2} + \frac{\omega_{k-1}}{K_4}+\omega_k\\
& \leq \frac{a_1}{K_4^k}+\frac{\omega_1}{K_4^{k-1}}+ \cdots+\frac{\omega_N}{K_2^{k-N}}+ \left(\frac{\omega_{N+1}}{K_4^{k-N-1}}+\cdots +\frac{\omega_{k-1}}{K_4}+\omega_k\right)\\
&\le (a_1+\omega_1+ .. .+\omega_N) \frac{1}{K_4^{k-N}} + \frac{1-(1/K_4)^{k-N}}{1-1/K_4} \epsilon
\to 0, \quad\mbox{as $k\to\infty$},
\end{align*}
with probability tending to one. Therefore, 
\begin{align*}
\Pr \left( \lim_{k \to \infty} \left\| \hat{\bbeta}_1^{(k)} - \hat{\bbeta}_1^{\circ} \right\|_2 = \mathbf{0} \right) = 1
\end{align*}
with probability tending to one, or equivalently
\begin{equation}
\label{eq1:result22}
\Pr(\hat{\bbeta}_1 = \hat{\bbeta}_1^{\circ}) = 1
\end{equation}
with probability tending to one. This proves (\ref{eq1:nzero_est}) and thus complete the proof of the theorem.
$\Box$ 

\subsection{Proof of Theorem \ref{th1:group}
.}
\label{group_proof}

{\bf{Proof:}} 
Under Conditions (C1) - (C6), by Theorem \ref{th1:oracle} we have that $\hat{\bbeta} = \displaystyle \lim_{k \to \infty} \hat{\bbeta}^{(k)}$, where 
\begin{align*}
\hat{\bbeta}^{(k+1)} = g(\hat{\bbeta}^{(k)}) = \mbox{arg} \min_{\bbeta}  \left\{-2l_n(\bbeta) + \lambda_n \sum_{j =1}^{p_n} \frac{I(\beta_j \neq 0) \beta_j^2}{ \left(\hat{\beta_j}^{(k)}\right)^2} \right\}.
\end{align*}
Note that
\begin{equation}
D(\hat{\bbeta}^{(k)})^{-1}\dot{l}_n(\hat{\bbeta}^{(k+1)}) = \lambda_n \hat{\bbeta}^{(k+1)}.\notag
\end{equation}
Therefore for any $l = i, j$ where $\hat{\beta}_i \neq 0$, $\hat{\beta}_j \neq 0$,
\begin{equation}
\label{derivi}
\hat{\beta}_l^{(k+1)} = \frac{ (\hat{\beta}_l^{(k)})^2}{\lambda_n} \dot{l}_{nl}(\hat{\bbeta}^{(k+1)}). \notag
\end{equation}
Letting $k \to \infty$, (\ref{derivi}), we have
\begin{equation}
\label{derivi2}
\hat{\beta}_l^{-1} = \frac{1}{\lambda_n} \dot{l}_{nl}(\hat{\bbeta}). \notag
\end{equation}
Let $\boldeta = X\bbeta$ and
\begin{equation}
\zeta(\eta_i) = \frac{\partial}{\partial \eta_i} l_n(\bbeta) = N_i(1) - \int_{0}^1 \frac{Y_i(s) \exp(\eta_i)}{\sum_{j=1}^n Y_{j}(s)\exp(\eta_j)} d\bar{N}(s) \hspace{.2in} i = 1, \ldots, n. \notag
\end{equation}
Then
\begin{equation}
\left| \zeta(\hat{\eta}_i) \right|   \leq \left| N_i(1) \right| + \left| \int_{0}^1 \frac{Y_i(s) \exp(\hat{\eta}_i)}{\sum_{j=1}^n Y_{j}(s)\exp(\hat{\eta}_j)} d\bar{N}(s) \right| \leq 1 + d_n \hspace{.2in} i = 1, \ldots, n, \notag
\end{equation}
 where $d_n = \sum_{i = 1}^n \delta_i$. Hence
\begin{equation}
\left\| \zeta(\hat{\boldeta}) \right\|_2 \leq  \left\| \boldsymbol{1} + d\boldsymbol{1} \right\|_2 = \sqrt{n(1+d)^2}. \notag
\end{equation}
Let $\mathbf{x}_{[,i]}$ denote the $i^{th}$ column of $X$. Since $X$ is assumed to be standardized, $\mathbf{x}_{[,i]}^T\mathbf{x}_{[,i]} = n - 1$ and $\mathbf{x}_{[,i]}^T\mathbf{x}_{[,j]} = (n-1)r_{ij}$, for all $i \neq j$ and where $r_{ij}$ is the sample correlation between $x_{[,i]}$ and $x_{[,j]}$. Since
\begin{equation}
\hat{\beta}_i^{-1} = \frac{1}{\lambda_n} \mathbf{x}_{[,i]}^T \zeta(\hat{\boldeta})  \notag
\hspace{.1in} \mbox{ and } \hspace{.1in}
\hat{\beta}_j^{-1} = \frac{1}{\lambda_n} \mathbf{x}_{[,j]}^T \zeta(\hat{\boldeta}), \notag
\end{equation}
we have   
\begin{align}
\left|\hat{\beta}_i^{-1} - \hat{\beta}_j^{-1} \right| & = \left|  \frac{1}{\lambda_n} \mathbf{x}_{[,i]}^T  \zeta(\hat{\boldeta})  - \frac{1}{\lambda_n} \mathbf{x}_{[,j]}^T \zeta(\hat{\boldeta}) \right|  \notag \\
& = \left| \frac{1}{\lambda_n} (\mathbf{x}_{[,i]} - \mathbf{x}_{[,j]})^T \zeta(\hat{\boldeta}) \right| \notag  \\
& \leq \frac{1}{\lambda_n} \left\| (\mathbf{x}_{[,i]} - \mathbf{x}_{[,j]}) \right\| \left\| \zeta(\hat{\boldeta}) \right\| \notag \\
& \leq \frac{1}{\lambda_n} \sqrt{2\{ (n-1) - (n-1)r_{ij}\}} \sqrt{n(1+d)^2}  \notag
\end{align}
for any $\hat{\beta}_i \neq 0$ and $\hat{\beta}_j \neq 0$. $\Box$

\subsection{Proof of Theorem \ref{th1:sjs}.}
\label{sjs_proof}

{\bf{Proof:}} 
Part (a) is a direct consequence of
Theorem 2 of \cite{yang2016feature} and part (b) is a consequence of part (a) and Theorem \ref{th1:oracle}.
 $\Box$

\subsection{Simulation results for ultrahigh dimensional data}
\label{ap1:hd_sim}

This section presents a simulation to illustrate the performance of our two-stage estimator SJS-CoxBAR described in Section \ref{s1:ultrahigh}
 in ultrahigh dimensional settings where $p_n$ is much larger than $n$. 
We generated data similar to Section \ref{s1:model_sim} with $n = 300$, $p_n = 2500$, $5000$, and 100 replications. For each simulated dataset, the sure joint screening method of \cite{yang2016feature} was initially used to choose a sub-model of size  $m  = \lfloor \frac{n}{\ln(n)} \rfloor = 52$, where$\lfloor \cdot \rfloor$ is the floor function. Using the sub-model obtained from sure joint screening, we compared the performance of hard thresholding (SJS-HARD), LASSO (SJS-LASSO), SCAD (SJS-SCAD), adaptive LASSO (SJS-ALASSO) and CoxBAR (SJS-$L_0$-CoxBAR, SJS-BIC-CoxBAR, SJS-cBIC-CoxBAR) on the screened model. BIC score minimization was used to select the optimal tuning parameter for SJS-HARD, SJS-LASSO, SJS-SCAD, SJS-ALASSO, and SJS-$L_0$-CoxBAR; while fixing $\lambda_n = \ln(n)$ and $\lambda_n = \ln(d_n)$ was used for SJS-BIC-CoxBAR and SJS-cBIC-CoxBAR, respectively. Similarly,  SJS-$L_0$-CoxBAR, SJS-BIC-CoxBAR,  SJS-cBIC-CoxBAR, and SJS-ALASSO had $\xi_n = 1$. As suggested by a referee, we also performed hard thresholding of the  Cox ridge estimator. We chose two values of the ridge tuning parameter, RIDGE$_1$ ($\xi_n = 50$) and RIDGE$_2$ ($\xi_n = 60$), and used BIC minimization to produce the hard-thresholded Cox ridge estimator. 
The simulation  results  are reported in Table A.1.  

Both ridge hard-thresholding methods have higher average numbers of false negatives compared to the two-step screening methods. We can also observe that there is a slight tradeoff between the number of false negatives and false positives depending on the tuning parameter used for the Cox ridge regression, which may suggest that the hard-thresholded ridge estimator is sensitive to the choice of $\xi_n$. Although comparable to each other, as in Section \ref{s1:model_sim}, the data-driven tuning parameter selected methods select an overwhelming number of false positives which, as a consequence, inflates the estimation bias.  Interestingly, both SJS-BIC-CoxBAR and SJS-cBIC-CoxBAR have much lower estimation bias and average number of false positives with slightly more false negatives when compared to the other procedures. Finally, we observe that although SJS-HARD, SJS-ALASSO, and SJS-$L_0$-CoxBAR generally have the smallest BIC scores, these methods tend to have substantially more false positives than BIC-CoxBAR and cBIC-CoxBAR.

\begin{table}
\caption{(High dimensional, moderate sample size) Simulated estimation and variable selection performance of SJS-$L_0$-CoxBAR, SJS-HARD, SJS-LASSO, SJS-SCAD, SJS-ALASSO, RIDGE$_1$ and RIDGE$_2$
(SJS-BIC-CoxBAR and SJS-cBIC-CoxBAR denote CoxBAR with $\lambda_n = \ln(n)$ and $\lambda_n = \ln(d_n)$ respectively; RIDGE$_1$ and RIDGE$_2$ denote hard thresholding the Cox ridge estimator of the original data with $\xi_n = 50$ and  $\xi_n = 60$, respectively;
SSB = sum squared bias; 
$P_j$ = probability that $\beta_{0j}$ is correctly identified;
FN = mean number of false positives; FP = mean number of false negatives; 
TM = probability that the selected model is equal to the true model;
AIC = AIC score;
BIC = BIC score;
ACR = average number of correctly ranked non-zero covariates; 
Each entry is based on 100 Monte Carlo samples with censoring rate = $20\%$) } 
\label{tab1:s34_results}
\centering
\setlength{\tabcolsep}{2.5pt}
\fbox{%
\begin{tabular}{rrrrrrrrrrrrrr}
$n = 300$, $p_n = 2500$   & SSB & $P_1$ & $P_3$ & $P_5$ & $P_6$ & $P_9$ & $P_{10}$ & FN & FP &  TM & AIC & BIC & ACR \\ 
  \hline
RIDGE$_1$ & 0.33 & 0.03 & 0.36 & 0.99 & 1.00 & 1.00 & 1.00 & 1.62 & 0.59 & 0.00 & 2100.56 & 2118.96 & 3.14 \\ 
  RIDGE$_2$ & 0.39 & 0.03 & 0.42 & 1.00 & 1.00 & 1.00 & 1.00 & 1.55 & 0.79 & 0.00 & 2106.04 & 2125.45 & 3.24 \\
  SJS-BIC-CoxBAR & 0.12 & 0.27 & 0.92 & 1.00 & 1.00 & 1.00 & 1.00 & 0.81 & 0.83 & 0.12 & 2051.48 & 2073.78 & 3.90 \\ 
  SJS-cBIC-CoxBAR & 0.12 & 0.29 & 0.92 & 1.00 & 1.00 & 1.00 & 1.00 & 0.79 & 1.11 & 0.11 & 2048.64 & 2072.04 & 3.93 \\ 
  SJS-CoxBAR & 1.89 & 0.48 & 0.93 & 1.00 & 1.00 & 1.00 & 1.00 & 0.59 & 24.40 & 0.00 & 1906.83 & 2017.24 & 3.34 \\ 
  SJS-HARD & 3.28 & 0.48 & 0.94 & 1.00 & 1.00 & 1.00 & 1.00 & 0.58 & 27.47 & 0.00 & 1906.83 & 2028.65 & 3.05 \\ 
  SJS-LASSO & 2.65 & 0.51 & 0.96 & 1.00 & 1.00 & 1.00 & 1.00 & 0.53 & 40.22 & 0.00 & 1914.97 & 2084.20 & 3.19 \\ 
  SJS-SCAD & 3.02 & 0.48 & 0.95 & 1.00 & 1.00 & 1.00 & 1.00 & 0.57 & 35.90 & 0.00 & 1903.49 & 2056.57 & 3.10 \\ 
  SJS-ALASSO & 2.16 & 0.48 & 0.94 & 1.00 & 1.00 & 1.00 & 1.00 & 0.58 & 31.89 & 0.00 & 1913.52 & 2051.71 & 3.27 \\   	         
   \hline
    \hline
$n = 300$, $p_n = 5000$ & & & & & & & & & & & & \\
  \hline
RIDGE$_1$ & 0.62 & 0.05 & 0.64 & 1.00 & 1.00 & 1.00 & 1.00 & 1.31 & 2.36 & 0.00 & 2118.44 & 2144.55 & 3.78 \\ 
  RIDGE$_2$ & 0.68 & 0.05 & 0.65 & 1.00 & 1.00 & 1.00 & 1.00 & 1.30 & 2.64 & 0.00 & 2125.60 & 2152.78 & 3.80 \\ 
 SJS-BIC-CoxBAR & 0.15 & 0.23 & 0.93 & 0.99 & 1.00 & 1.00 & 1.00 & 0.85 & 1.51 & 0.08 & 2038.38 & 2063.05 & 3.75 \\ 
  SJS-cBIC-CoxBAR & 0.16 & 0.23 & 0.93 & 0.99 & 1.00 & 1.00 & 1.00 & 0.85 & 1.94 & 0.04 & 2034.24 & 2060.50 & 3.73 \\ 
  SJS-CoxBAR & 1.87 & 0.33 & 0.95 & 0.99 & 1.00 & 1.00 & 1.00 & 0.73 & 22.85 & 0.00 & 1899.74 & 2003.89 & 3.44 \\ 
  SJS-HARD & 3.08 & 0.31 & 0.96 & 0.99 & 1.00 & 1.00 & 1.00 & 0.74 & 25.70 & 0.00 & 1898.81 & 2013.48 & 3.32 \\ 
  SJS-LASSO & 2.35 & 0.39 & 0.96 & 0.99 & 1.00 & 1.00 & 1.00 & 0.66 & 38.57 & 0.00 & 1913.29 & 2075.92 & 3.46 \\ 
  SJS-SCAD & 2.89 & 0.36 & 0.96 & 0.99 & 1.00 & 1.00 & 1.00 & 0.69 & 35.04 & 0.01 & 1895.52 & 2044.96 & 3.45 \\ 
  SJS-ALASSO & 1.93 & 0.35 & 0.96 & 0.99 & 1.00 & 1.00 & 1.00 & 0.70 & 30.19 & 0.00 & 1906.38 & 2037.82 & 3.51 \\ 
      \hline
\end{tabular}
}
\end{table}

\subsection{Diffuse Large-B-Cell lymphoma data}
\label{s1:blca}
For an application of SJS-CoxBAR in the ultrahigh dimensional setting, we analyze a microarray diffuse large-B-cell lymphoma dataset \citep{rosenwald2002use}. The dataset consists of $240$ DLBCL patients and $7399$ cDNA microarray expressions. The censoring rate was around $43\%$. Interest lies in understanding and identifying the genetics markers that may impact survival. Due to the large number of covariates and relatively small sample size, variable screening is an important step to reducing the dimensionality of the problem. 

Our analysis was similar to \cite{yang2016feature}. The  covariates were standardized to have mean zero and variance one and we remove the 5 patients whose observed survival times were close to 0. To reduce the number of genes in the analysis, sure joint screening was used to obtain a reduced model with $43$ genes. These genes were identified in \cite{yang2016feature} who then performed LASSO and SCAD on the reduced model. The optimal tuning parameter for LASSO and SCAD were found using BIC score minimization. We apply our CoxBAR method with $\lambda_n = \ln(n)$ (SJS-BIC-CoxBAR), $\lambda_n = \ln(d_n)$ (SJS-cBIC-CoxBAR), and $\lambda_n$ found using BIC score minimization (SJS-$L_0$-CoxBAR) to the same $43$ genes and compare our results to the LASSO and SCAD results reported in \cite{yang2016feature}. As with the other numerical results, we fix $\xi_n = 1$. These results are provided in Table \ref{tab1:blca}.

We see that the ordering of the BIC scores from Table \ref{tab1:blca} are reflective of the ordering in Table \ref{tab1:s34_results}, with SJS-$L_0$-CoxBAR having the smallest BIC score while both SJS-BIC-CoxBAR and SJS-cBIC-CoxBAR have larger BIC values. All three data driven methods also select far more variables than SJS-BIC-CoxBAR and SJS-cBIC-CoxBAR, a similar observation to our simulation studies. Finally, the genes identified by SJS-BIC-CoxBAR and SJS-cBIC-CoxBAR are a subset of those identified by SJS-$L_0$-CoxBAR, SJS-SCAD, and SJS-LASSO.

\begin{table}
\caption{(BLCA data)  Comparison of SJS-LASSO, SJS-SCAD, SJS-ALASSO, 
and SJS-CoxBAR
for the BLCA data. (BIC-CoxBAR and cBIC-CoxBAR denote CoxBAR with $\lambda_n = \ln(n)$ and $\lambda_n = \ln(d_n)$ respectively;
 SJS-LASSO and SJS-SCAD results are from \cite{yang2016feature})}
\centering
\label{tab1:blca}
\fbox{%
\begin{tabular}{r|rrrr}
 Method & Log-partial likelihood & $\#$ Selected & BIC Score\\ 
 \hline
 SJS-SCAD & -546.1902 & 30 & 1256.168 \\
 SJS-LASSO & -542.9862 & 36 & 1282.518 \\
SJS-$L_0$-CoxBAR & -558.9954 & 
20 & 1227.182 \\
 SJS-BIC-CoxBAR & -624.1901 & 5 & 1275.678 \\
 SJS-cBIC-CoxBAR & -607.2283 & 7 & 1264.964 \\
\end{tabular}
}
\end{table}

\end{document}